\documentclass[fleqn,usenatbib]{mnras}

\usepackage{newtxtext,newtxmath}

\usepackage[T1]{fontenc}
\DeclareRobustCommand{\VAN}[3]{#2}
\let\VANthebibliography\thebibliography
\def\thebibliography{\DeclareRobustCommand{\VAN}[3]{##3}\VANthebibliography}

\usepackage{graphicx}	
\usepackage{amsmath}	
\usepackage{subcaption}
\usepackage[normalem]{ulem}

\title[Galaxy interactions in the most luminous AGN]{Interacting galaxies in the IllustrisTNG simulations - VII: The connection between the most luminous active galactic nuclei and galaxy interactions}

\author[S. Byrne-Mamahit et al.]{Shoshannah Byrne-Mamahit,$^{1}$\thanks{E-mail: sjbyrnem@uvic.ca}
David R. Patton,$^{2}$
Sara L. Ellison,$^{1}$
Robert Bickley,$^{1}$
\newauthor Leonardo Ferreira,$^{1}$
Maan Hani,$^{1,3}$
Salvatore Quai,$^{4,5}$
Scott Wilkinson$^{1}$
\\
$^{1}$ Department of Physics \& Astronomy, University of Victoria, 3800 Finnerty Road, Victoria, British Columbia, V8P 5C2, Canada\\
$^{2}$ Department of Physics \& Astronomy, Trent University, 1600 West Bank Drive, Peterborough, Ontario, K9L 0G2, Canada\\
$^{3}$ Department of Physics \& Astronomy, McMaster University, 1280 Main Street W, Hamilton, Ontario, L8S 4K1, Canada\\
$^{4}$ Dipartimento di Fisica e Astronomia “Augusto Righi,” Universit\`{a} degli Studi di Bologna, Via Gobetti 93/2, I-40129 Bologna, Italy \\
$^{5}$ INAF-Osservatorio di Astrofisica e Scienze dello Spazio di Bologna, Via Gobetti 93/3, I-40129 Bologna, Italy}

\date{Accepted XXX. Received YYY; in original form ZZZ}

\pubyear{2023}

\begin{document}
\label{firstpage}
\pagerange{\pageref{firstpage}--\pageref{lastpage}}
\maketitle

\begin{abstract}
We investigate the connection between the most luminous active galactic nuclei (AGN), galaxy pairs, and post-mergers in the IllustrisTNG simulation. We select galaxy pairs and post-mergers with a mass ratio between 1:10 $< \mu <$ 1:1 and a redshift between $0<z<1$. We compare the incidence of luminous AGN in pairs with matched non-pair controls, finding that AGN with luminosity $L_{\mathrm{bol}}>10^{44}$ erg/s occur in $\sim $26\% of paired galaxies with a companion within 20 kpc, compared with $\sim $14\% in controls (a fractional excess of $\sim$2). The enhancement of AGN in galaxy pairs is luminosity dependent, with the highest excess (a factor of $\sim6\pm2$ at the closest separations) for AGN with $L_{\mathrm{bol}}>10^{45}$ erg/s. Additionally, pairs exhibit a modest yet statistically significant excess of luminous AGN up to $\sim$150 kpc in separation. For pairs which merge between $0<z<1$, AGN fractions are elevated between 1.5 Gyr before and after coalescence, with the highest excess closest to coalescence. Our results indicate that pre-coalescence interactions drive excesses of luminous AGN, but that luminous AGN in galaxy pairs are not ubiquitous. Finally, we investigate what fraction of AGN can be associated with an interaction (either having a companion within 100 kpc or a merger within the last 500 Myr). For AGN with $L_{\mathrm{bol}}>10^{45}$ erg/s, $\sim$55\% are interacting, compared with a 30\% interaction fraction in AGN with $10^{44}<L_{\mathrm{bol}}<10^{44.5}$ erg/s. Our results support a picture in which interactions play a dominant role in (but are not the sole cause of) triggering the most luminous AGN. 
\end{abstract}

\begin{keywords}
galaxies: interactions -- galaxies: active -- galaxies: evolution
\end{keywords}



\section{Introduction}
\label{sec:introduction}

Galaxy mergers play a critical role in the growth and evolution of galaxies. Mergers are a ubiquitous process within the cosmological model of hierarchical assembly \citep[][]{WhiteReese1978,Springel2005}{}{}, as well as a morphologically tranformative process through the dynamic rearrangement of the galaxy's stellar matter \citep[][]{Toomre1972}{}{}. The influence of merger events on the star formation of galaxies has also been well established, where pair interactions and mergers have been demonstrated to fuel star formation rate enhancements in both idealized simulations \citep[][]{Hernquist1989a,Barnes1991,Mihos1992,DiMatteo2007}{}{} and observational studies \citep[][]{Barton2000,Ellison2008,Woods2010,Scudder2012,Patton2013,Ellison2013,Thorp2019,Bickley2022}, although the strength of such star formation enhancements will vary with different galaxy and orbit properties \citep[][]{Woods2006,Woods2007,Cox2008,Lotz2008,Knapen2015,Moreno2015,Cao2016,Brown2023}{}{}. Star formation rate enhancements are made possible by the transport of gas towards the galactic centre in interactions \citep[][]{Capelo2016,Blumenthal2018}{}{}, where it may also enhance the accretion of gas onto supermassive black holes (SMBHs) fueling active galactic nuclei (AGN), as shown in idealized simulations of galaxy mergers \citep[][]{DiMatteo2005,Springel2005}{}{}. 

In fact, a theoretical relationship between galaxy mergers, enhanced SMBH accretion, and AGN feedback has been demonstrated to naturally reproduce the incidence of highly luminous AGN in the Universe \citep[][]{Hopkins2008}{}{}, and would additionally provide a causal connection between the growth of the SMBH and the larger scale galaxy properties as seen in observations \citep[][]{Ferrarese2000,Kormendy2001,Tremaine2002}{}{}. However, such correlations have been shown to be reproducible without a causal coupling of the SMBH and galaxy bulge \citep[][]{Jahnke2011}{}{}. 

Despite the well founded theoretical links between mergers and AGN triggering, there exists considerable tension in the observational literature. We discuss below the findings from previous studies of the merger-AGN connection, which may be broadly separated into two categories: those that investigate occurrences of AGN triggering in the galaxy pair and post-merger populations, and those that investigate the occurrences of pairs and post-mergers in the AGN population.

Among the former, many studies find that pair and post-merger populations host an excess of AGN when compared with non-interacting controls \citep[][]{Alonso2007,Ellison2011,Ellison2013,Goulding2018,Bickley2023,Li2023}{}{}. In addition, a number of studies also demonstrate the highest excesses in mid-IR selected AGN \citep[][\citealt{Dougherty2023} for the most obscured AGN]{Satyapal2014,Weston2017,Secrest2020}{}{}, which may be due to the enhancement of dust obscuration in merger triggered AGN. Despite this, the overall fraction of AGN in the pair and post-merger populations remains low, demonstrating that AGN triggering in pairs and post-mergers is rare, and studies of high redshift interacting galaxies often find no statistically significant excess of AGN in pairs and post-mergers \citep[][]{Shah2020,Silva2021,Dougherty2023}{}{}, suggesting that the merger-AGN connection may be less significant with increasing redshift. 

The alternative approach is to assess whether a significant fraction of the AGN population is associated with pairs or post-mergers. While some observational studies find that a high fraction of AGN reside in morphologically disturbed hosts, these are usually in the most luminous \citep[][]{Bennert2008,Urrutia2008,Bessiere2012,Glikman2015,Hong2015,Pierce2022}{}{} or most obscured \citep[][]{Kocevski2015,Ellison2019,Fan2016,Gao2020}{}{} AGN samples, and are contradicted by studies that do not find a significant fraction of post-mergers or interacting pairs in AGN hosts \citep[][\citealt{Villforth2023} for a comprehensive literature review]{Schawinski2011,Schawinski2012,Hewlett2017,Villforth2014}{}{}. In addition, there are contradicting results as to whether there is an excess of interactions over non-AGN controls \citep[][]{Koss2010,Kocevski2015,Rosario2015,Ellison2019,Gao2020,Marian2020}{}{}, or whether AGN hosts have a normal incidence of interacting systems \citep[][]{Cisternas2011,Kocevski2012,Bohm2013,Mechtley2016,Villforth2017,Marian2019}{}{}.

An additional facet to the connection between mergers and AGN is the dynamical difference between pair phase interactions, coalescence, and the post-merger phase. Simulations suggest that starbursts and AGN activity peak at coalescence and are weaker during the interacting pair phase \citep[][]{DiMatteo2005}{}{}, and similarly observations have found a higher excess of AGN in post-mergers than in pairs \citep[][]{Ellison2013,Bickley2023}{}{}. Despite this, recent work with numerical simulations demonstrate that pair interactions can fuel quasar-like AGN events \citep[][]{Prieto2021}{}{}, and that a significant (although still sub-dominant) fraction of AGN reside in close pairs \citep[][]{Bhowmick2020}{}{}.

In contrast to the abundance of observational studies investigating the connection between mergers and luminous AGN, relatively few theoretical studies exist which investigate AGN triggering in a large and representative sample of mergers from cosmological simulations \citep[][]{Steinborn2018,Bhowmick2020,McAlpine2020}{}{}, with many studies investigating the connection using simulation suites with a limited number of merger configurations \citep[][]{DiMatteo2005,Hopkins2005,Capelo2015,Volonteri20151,Blecha2018}{}{}. In the work presented here, we make use of the IllustrisTNG cosmological simulation (hereafter TNG) \citep[][]{Nelson2017,Naiman2018,Pillepich20171,Springel2017,Mariancci2018,Nelson2019b}{}{}. We use the TNG simulation to study a diverse and representative sample of pairs and post-mergers. Additionally, TNG allows us to study complete and pure samples of post-mergers, a significant challenge in observational studies (see \citealt{Wilkinson2024} for a comprehensive analysis of post-merger observability). The research presented here is part of an ongoing series which investigates the merger sequence in simulated post-merger and pair galaxies from TNG: Interacting Galaxies in the IllustrisTNG Simulations. In this series, we investigated star formation rate enhancements in the pair \citep[][]{Patton2020,Brown2023}{}{} and post-merger \citep[][]{Hani2020}{}{} phase, the enhancement of SMBH accretion rates \citep[][]{ByrneMamahit2023}{}{} and the incidence of rapid quenching \citep[][]{Quai2021}{}{} in post-merger galaxies, and characterized the orbits of TNG pair galaxies \citep[][submitted]{Patton2023}{}{}.

In \cite{ByrneMamahit2023} we investigated the enhancement of SMBH accretion rates in TNG post-mergers compared with non-interacting controls. We found a luminosity dependence in the fraction of AGN in post-merger galaxies, and that the most luminous AGN are more commonly found in post-mergers by a factor of 3-4 when compared with matched controls. Despite this, our findings show that post-mergers make up a subdominant percentage of even the most luminous AGN in TNG (10-20\% of $L_{\mathrm{bol}}\geq 10^{45}$ erg/s). We demonstrate the same behaviour in post-mergers from both the EAGLE and Illustris simulations in \cite{Quai2023}. However, none of these studies additionally investigate the role of pair phase interactions, which is the subject of the work presented here.

The structure of this paper is as follows. In Section \ref{sec:methods}, we summarize the relevant SMBH models from the IllustrisTNG simulation. In addition, we outline our pair and post-merger selection methodology, control matching methodology, and AGN luminosity calculations. In Section \ref{sec:AGNinPairsAndPMs}, we study the fraction of TNG pairs which host luminous AGN, quantify the fractional excess over matched non-pair controls, and investigate how the AGN fraction progresses along the merger sequence. In Section \ref{sec:IntFracOfAGN}, we present the fraction of interactions in our AGN sample as well as the excess of interactions over matched non-AGN controls. We discuss how the excess of interactions in AGN depend on properties such as redshift and host galaxy properties, and discuss comparisons with observational studies in Section \ref{sec:discussion}. Our major findings are summarized in Section \ref{sec:conclusion}.

\section{Methods}
\label{sec:methods}

\subsection{Overview of AGN in IllustrisTNG}
\label{subsec:IllustrisTNG}

In the work presented here, we use the IllustrisTNG cosmological magneto-hydrodynamic galaxy formation simulation \citep[][]{Nelson2017,Pillepich20171,Springel2017,Mariancci2018,Naiman2018,Nelson2019b,Nelson2019,Pillepich2019}. There are three different volume and resolution runs of the fiducial TNG model, of increasing resolution and decreasing volume: TNG300-1, TNG100-1, and TNG50-1. In order to maximize the statistics of our samples while maintaining a sufficient resolution to resolve intermediate mass galaxies, we use the intermediate volume and resolution run, TNG100-1, which has a $(110.7 \, \mathrm{Mpc})^3$ volume, a baryonic resolution of $1.4 \times 10^6 \, \mathrm{M_{\odot}}$, and a dark matter resolution of $7.5 \times 10^6 \, \mathrm{M_{\odot}}$. As our research concerns the implementation of SMBHs and AGN in TNG, we summarize the details of the SMBH models below, and refer the reader to \cite{Weinberger2017} and \cite{Pillepich2018} for a comprehensive description of the physics models. TNG, as well as all other galaxy formation models, does not directly simulate many SMBH physical processes, such as SMBH formation, accretion disks and black hole mergers. It is therefore necessary to implement parameterized models and recipes in order to simulate the SMBH and AGN processes, called `sub-grid' models, which we briefly describe.

SMBHs are initialized as a seed particle, with a mass of $M_{BH} = 8 \times 10^5 \, \mathrm{h^{-1} M_{\odot}}$, at the centre of halos with a mass of at least $5 \times 10^{10} \, \mathrm{h^{-1} M_{\odot}}$. After seeding, SMBHs can grow in mass through the accretion of material from the region surrounding the black hole. The mass accretion rate for SMBHs is calculated using a Bondi-Hoyle-Lyttleton model for spherically symmetric accretion \citep[][]{Hoyle1939,Bondi1944,Bondi1952}{}{},
\begin{equation}
    \dot M_{Bondi} = \frac{4 \pi G^2 M_{BH}^2 \rho}{c_s^3},
    \label{eq:Bondi}
\end{equation}
where $G$ is the gravitational constant, $M_{BH}$ is the black hole mass, and $\rho$ and $c_s$ are the gas density and sound speed sampled in a kernel-weighted sphere enclosing $\sim $128 cells, centred on the SMBH. In addition, SMBH accretion is limited to the Eddington rate, 
\begin{equation}
    \dot M_{Edd} = \frac{4 \pi G M_{BH} m_p}{\epsilon_r \sigma_T c},
    \label{eq:Edd}
\end{equation}
where $m_p$ is the proton mass, $\epsilon_r$ is the radiative accretion efficiency, $\sigma_T$ is the Thompson cross-section, and $c$ is the vacuum speed of light. 

When the distance between SMBHs is smaller than the kernel-weighted sphere over which the accretion rate is calculated, the black holes are merged. SMBH mergers often occur promptly during the coalescence of galaxies \citep[][]{Bahe2022}{}{}, as SMBHs are re-positioned to their local gravitational potential minima to avoid the wandering of SMBHs away from the halo centre. Additionally, the SMBH repositioning can have side-effects such as SMBH stripping, when a galaxy loses its SMBH to a companion \citep[][]{Borrow2023}{}{}.

In TNG, SMBH feedback is implemented using a dual feedback mode model. The SMBH feedback mode depends on the SMBH mass and the SMBH accretion rate, according to the following prescription: high vs. low accretion rates are defined by the following relation,
\begin{equation}
    \dot M_{BH}^{high} \geq \chi \dot M_{Edd} \hspace{0.5cm},\hspace{0.5cm} \chi = min\bigg[\chi_0 \bigg(\frac{M_{BH}}{10^8 M_{\odot}}\bigg)^{\beta},0.1\bigg].
    \label{eq:chi}
\end{equation}
$\chi_0$ and $\beta$ are simulation parameters tuned to 0.002 and 2 respectively, and $\chi$ is capped to 0.1, following observational constraints set by X-ray binaries \citep[][]{Dunn2010}{}{}. Energy from AGN feedback is coupled to the medium surrounding the SMBH differently according to whether the SMBH has a high or low accretion rate. At high accretion rates ($\dot M_{BH} \geq \chi \dot M_{Edd}$) a fraction of the accreted mass energy is instantaneously returned to the galaxy as thermal energy, heating up the gas particles surrounding the black hole. At low accretion rates ($\dot M_{BH} < \chi \dot M_{Edd}$), the accreted mass energy is stored in a reservoir and then released into the gas surrounding the SMBH region in the form of kinetic energy, in randomized directions away from the black hole. For additional details on the specific calibrations used in the AGN feedback models, we refer the reader to \cite{Weinberger2017}.

\subsection{Galaxy pair and post-merger selection}
\label{subsec:IDPairs_and_mergers}

We select our galaxy sample from TNG100-1 by applying the following global selection criteria. First, we exclude galaxies at $z>1$. We limit our redshift range primarily due to the increased frequency of mergers and interactions at higher redshift, which exacerbates issues such as numerical stripping and subhalo switching \citep[][]{Rodriguez-Gomez2015}, and make robust merger identification more difficult. We additionally reject any galaxies with a \texttt{SubhaloFlag=False}, corresponding to galaxies which meet criteria suggesting that they may be of non-cosmological origin (for example, a spurious group of stellar particles that fragment from a larger galaxy). Finally, we apply a minimum stellar mass cut of $10^{9} M_{\odot}$, using the total stellar mass assigned to the subhalo, which is roughly equivalent to $\sim$1000 particles. The minimum stellar mass cut ensures the selected galaxies are well resolved, allowing for robust determination of properties such as the merger stellar mass ratio. Since the work presented here concerns interacting galaxies, we apply the minimum mass limit according to each subhalo's maximum total stellar mass in the last 500 Myr, in order to correct for numerical stripping\footnote{As galaxies interact at close separations, the numerical algorithms which assign constituent particles into distinct subhalos can `strip' mass from one of the galaxies and assign it to the nearby companion.} in interacting pairs \citep[][]{Patton2020}{}{}. Applying the above global constraints selects a parent sample of 1,021,226 TNG galaxies (an average of $\sim$ 20,000 galaxies per simulation snapshot). From this parent sample, we select two distinct samples; first a sample of pre-coalescence galaxy pairs and second a sample of post-mergers.

\begin{figure}
	\includegraphics[width=\columnwidth]{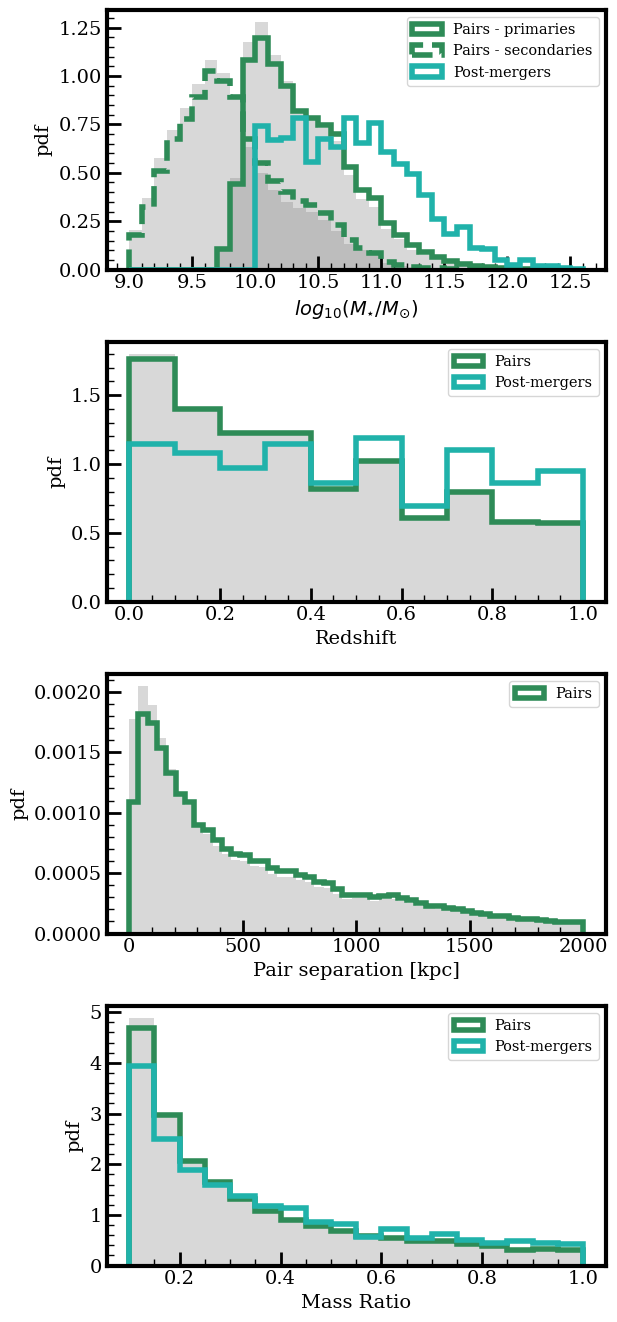}
    \caption{Properties of the pair and post-merger galaxies selected from TNG100-1. The grey histograms show the pair sample distributions before the pairwise removal of galaxies with their SMBH stripped. The solid and dashed green lines show the distribution of our final pairs sample (159,770 pairs of separations up to 2 Mpc), consisting of paired galaxies of masses greater than $10^{9}M_{\odot}$ individually and $10^{10}M_{\odot}$ combined. The solid green line shows the distribution for the more massive galaxy in each pair (primary), and the dashed green line shows the distribution for the less massive galaxy in each pair (secondary). The solid teal line shows the distribution for our sample of 2741 post-mergers selected in the snapshot immediately after coalescence, with a total stellar mass greater than $10^{10}M_{\odot}$. The top panel shows the stellar mass distribution for the post-mergers and pairs (split into primaries and secondaries). The panels below show the redshift distribution for post-mergers and pairs, and the distribution of pairs in bins of pair separation. The bottom panel shows the pair and post-merger mass ratios.}
    \label{fig:PairSample}
\end{figure}

We identify pairs from the parent sample using a methodology similar to that in \cite{Patton2020}. For every galaxy in the parent sample, we compute the 3D separation between galaxies and their closest companion (up to a distance of 2 Mpc) with a mass of at least 1:10 of the host mass. Once again, to correct for recent numerical stripping, the pair mass ratio is calculated using each galaxy's maximum mass in the previous 500 Myr.

Next, we identify post-mergers using the merger trees created by the \textsc{Sublink} algorithm \citep[][]{Rodriguez-Gomez2015}, which associate each galaxy in TNG with progenitor and/or and descendant galaxies. We begin by selecting all mergers, identified as nodes in the merger trees. The post-merger nodes occur when two distinct subhalos are associated with the same descendant subhalo in the following snapshot, allowing us to determine the time of a merger to within the snapshot temporal resolution, $\sim$160 Myr for our redshift range in TNG100-1. We first select all nodes in which the subhalos meet the minimum mass and SubhaloFlag criteria. We then apply the following additional conditions to select our final fiducial merger sample:

\begin{itemize}
    \item Mass Ratio: Although both the pairs and post-mergers in our sample have mass ratios within a factor of 1:10, we use a different mass ratio calculation for the two samples due to the significant mass stripping and mass exchange incurred in the pair phase and final stages of coalescence. We apply a modified version of the methodology from \cite{Rodriguez-Gomez2015}. Following \cite{Rodriguez-Gomez2015}, we calculate the mass ratio in the snapshot where the less massive progenitor has a maximum stellar mass. We apply an additional requirement that the maximum mass cannot occur while the merging galaxies are within 50 kpc, to ensure that the mass ratio is not calculated after significant (numerical or physical) mass exchange between the merging galaxies. If the maximum mass occurs at close separation, we calculate the mass ratio at the next snapshot of maximum mass. We additionally note that the minimum separation requirement indirectly reduces the chance that the mass ratio is measured after strong star formation rate enhancements that occur near coalescence, which, in combination with numerical stripping, can significantly reduce the mass ratio in the final snapshots before the merger event. 
    \item Spurious subhalo branches: We reject post-mergers in the case that the next progenitor's \textsc{Sublink} history is fewer than 3 snapshots long. We apply this constraint to avoid detecting mergers between a galaxy and a companion with an unreliable mass history.
    \item Isolated post-mergers: In order to isolate the difference in AGN triggering between the pair and post-merger phase, we restrict our post-merger sample to those without a pair companion within 100 kpc.
\end{itemize}

In addition to the minimum stellar mass cut applied globally to our TNG100-1 galaxy sample, we apply the following mass criteria in order to select complementary samples of pairs and post-mergers. Post-mergers are selected with a minimum total stellar mass of $10^{10}M_{\odot}$. The minimum mass cut combined with the minimum post-merger mass ratio of 1:10 ensures the progenitors of the post-merger have a stellar mass of at least $10^{9}M_{\odot}$. We select pairs where the individual galaxies have a minimum stellar mass of at least $10^{9}M_{\odot}$ and the pair has a combined stellar mass of at least $10^{10}M_{\odot}$. In the event that a galaxy is in multiple pairs which meet all of the above criteria (for example, two galaxies can share a closest companion) we only select the pair with the closer pair separation. We identify 177,473 pairs with separations up to 2 Mpc which meet all the criteria outlined above. Therefore, out of the 1,021,226 TNG galaxies in our parent sample, we have 354,946 paired galaxies: 177,473 primaries (more massive galaxy in each pair) and 177,473 secondaries (less massive galaxy in each pair), and 2741 post-mergers (in the snapshot immediately after coalescence). We expect that the number of pairs significantly outnumbers the post-mergers for two reasons. Firstly, we identify post-mergers as single events in the simulation time, whereas the same pair is selected at multiple times and different pair separations throughout the simulation. The selection of late-stage post-mergers is discussed in Seciton \ref{subsec:controls}. Secondly, we note that not all pairs experience a merger within the simulation time. In particular, \citet[][submitted]{Patton2023}{}{} investigates the merging fraction for pairs at different 3D separations, finding a merging fraction of only 50\% (within the next 8 Gyrs) for pair separations of 200-500kpc.

Finally, we further reduce our pair sample by removing cases in which one of the galaxies in a pair has been stripped of the SMBH (see Section \ref{subsec:IllustrisTNG} for details on SMBH stripping). The removal is performed pairwise in order to ensure the sample remains balanced in the number of primaries and secondaries. Following the removal of pairs with stripped SMBHs, we retain 159,770 of 177,473 pairs, for a total of 319,540 paired galaxies in our final sample.

Figure \ref{fig:PairSample} shows the characteristics of the galaxies in the pairs and post-merger samples. In all panels, the grey histogram shows the distribution of the pairs before the removal of pairs with stripped SMBHs. The final pair sample is shown in the green lines. In the top panel, the paired galaxies are separated into primaries (solid green line) and secondaries (dashed green line), and the post-merger sample is shown in the solid teal line. In the rest of the panels, only one line (solid green) is shown for the pairs since the parameters are shared between the primaries and secondaries. The total stellar mass is shown in the top panel of the figure. Below the total stellar mass is the redshift distribution of the pairs and post-mergers, where we note a decreasing number of pairs with increasing redshift. The drop in galaxy numbers at higher redshift is a feature of our mass selection, as we find more galaxies meeting our minimum mass threshold at lower redshift. The next panel below the redshift is the distribution of galaxy pair separation which is only shown for the paired galaxies. Finally, the bottom panel shows the distribution of pair and post-merger mass ratio. A consequence of the removal of pairs with stripped SMBH mass is a biased incompleteness of the pair sample. When comparing the green lines to the grey histograms, we find a strong bias incurred in the sample with pair separation. We note that we remove more than half of the pairs within 0-20 kpc, and $\sim $25\% of pairs between 20-50 kpc. The higher incidence of SMBH stripping in close pairs is expected, as the close interactions will increase the chances that the SMBH is stripped away from one of the galaxies in the pair.

\subsection{Control Matching Methodology}
\label{subsec:controls}

Similar to the previous works in the series on interacting galaxies in IllustrisTNG, we once again employ control matching techniques in order to quantify the enhancement of properties, such as SMBH accretion rates, in galaxies along the merger sequence when compared with matched non-interacting controls. In the work presented here, we employ different control matching schemes for the pairs and post-mergers. Considering first the pairs, we apply a control matching procedure to our pair sample in order to construct a redshift, stellar mass, and environment matched control sample, following the methodology of \citet[][]{Patton2020}{}{}. To begin, for every pair galaxy we select a control pool of galaxies that match exactly in redshift (i.e. match in simulation snapshot). We note that redshift matching, in addition to removing the influence of redshift between the pair and control samples, also ensures no galaxy is ever control matched to its own progenitor or descendant. From the control pool, we select the best simultaneous match in stellar mass and environment. To characterize the environment, we first match the distance of the second closest companion with a mass ratio of at least 1:10 in the paired galaxy to the closest companion in the non-pair control. Since the distance to the second closest companion is dependent on the minimum mass of each paired galaxy, it is possible for the primary and secondary galaxy in a pair to have a different second closest companion. In this way, our control sample may still include close pairs in order to preserve the local environment of pair galaxies within the control sample. For example, the best `non-pair' match for a paired galaxy with 2 close companions at 50 kpc and 70 kpc would be a galaxy with one companion at 70 kpc. We additionally match on the number of galaxies within a 2 Mpc volume that have a stellar mass of at least $10^{9} M_{\odot}$. The single best matched control is found by maximizing the following,
\begin{equation}
    \prod_i \Big(1-\frac{\mathrm{abs}(x_i^{\mathrm{pair}}-x_i^{\mathrm{control}})}{\delta x_i}\Big)
    \label{eq:control}
\end{equation}
where \textit{x} would be the matching parameters: total stellar mass, nearest neighbour separation, and number of neighbours within 2 Mpc, and $\delta x$ is the error tolerance on each matched parameter. For each pair, we initialize an error tolerance for all matching parameters. If no control can be found within the initial error tolerances, the error tolerance is relaxed up to a maximum error tolerance. If no control can be found within the maximum error tolerances, both the unmatched galaxy and its pair companion are removed from the pair sample. For our pair matching scheme, we initialize the error tolerances to 0.1 dex in stellar mass and 10\% in environment variables, with maximum error tolerances of 0.3 dex in stellar mass and 50\% in environment variables.  

Out of the 159,770 pairs, 2724 are removed after unsuccessful control matching, for a total final matched sample of 314,092 paired galaxies. For the matched paired galaxies, the median offset ($x_{i}^{\mathrm{pair}}-x_{i}^{\mathrm{control}}$) in the matched parameters is 0.04 dex in total stellar mass, 20 kpc in distance to the next closest companion, and 0 in the number of neighbours. 

We apply a separate control matching procedure to our post-merger sample in order to construct a comparative non-merger control sample. Additionally, for each of our 2741 post-mergers (immediately after coalescence), we attempt to control match the post-mergers and their descendants for up to 2 Gyr post-coalescence, identifying new controls at each subsequent snapshot after coalescence, for a total of 59,537 post-merger descendants. We once again select from a control pool matched exactly in redshift, however we require that the control pool additionally consist of galaxies that have not had a merger event of mass ratio > 1:10 within the past 2 Gyr. From the control pool, we select the best matched control using the same methodology (and error tolerances) as the pairs: matching in total stellar mass, distance to the nearest neighbour, and the number of neighbours within 2 Mpc.

Out of the 59,537 post-mergers and post-merger descendants, 621 are removed from the sample after unsuccessful control matching. Once again, for the matched post-mergers and descendants, the median offset in the matched parameters is 0.01 dex in total stellar mass, 6 kpc in distance to the closest companion, and 0 in the number of neighbours. Therefore, our fiducial samples consist of 157,046 control matched pairs with separations up to 2 Mpc (or 314,092 paired galaxies), and 58,916 control matched post-mergers (within 2 Gyr of coalescence). 

\subsection{Identifying AGN}
\label{subsec:TNGAGN}

\begin{figure}
	\includegraphics[width=\columnwidth]{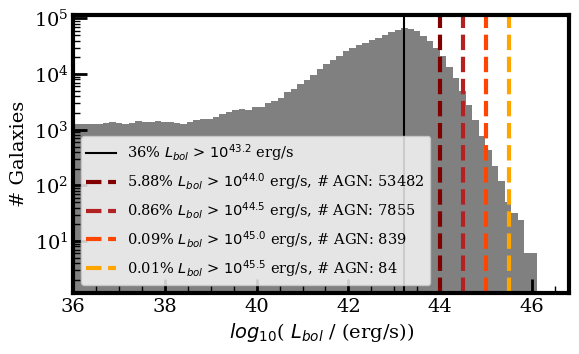}
    \caption{Selection of luminous AGN from the parent sample of TNG100-1 galaxies. The grey histogram shows the distribution of bolometric AGN luminosities, calculated using Eq. \ref{eq:Lbol}, for the parent sample of all galaxies above a total stellar mass of $10^{9} M_{\odot}$ and $z<1$. The solid black vertical line shows the most common AGN luminosity for galaxies from our parent sample, of $L_{\mathrm{bol}} \sim 10^{43.2}$ erg/s. The dashed lines define luminous AGN samples by their minimum $L_{\mathrm{bol}}$, colour coded from dark red to yellow, in order of increasing luminosity. The legend notes the number of galaxies that have AGN luminosities in excess of each $L_{\mathrm{bol}}$, and the percent relative to the entire parent sample.}
    \label{fig:AGNLums}
\end{figure}

We use the instantaneous SMBH accretion rates to assign a bolometric AGN luminosity to all of the selected galaxies from TNG100-1 meeting a minimum stellar mass of $10^9 M_{\odot}$ and $z<1$. We convert the SMBH accretion rates to bolometric luminosity using the following treatment.
\begin{equation}
    L_{\mathrm{bol}} = 0.1\times \dot M_{BH} \mathrm{c}^2,
    \label{eq:Lbol}
\end{equation}
where 0.1 reflects a 10\% efficiency in the conversion of accreted mass to radiative energy. The distribution of AGN luminosity for the entire parent sample is shown in the grey histogram of Figure \ref{fig:AGNLums}. In the figure, we note that the most common AGN luminosity in our parent sample is $L_{\mathrm{bol}}\sim 10^{43.2}$ erg/s, shown in the solid black line. Approximately 36\% of the parent sample have AGN luminosities in excess of this limit. We select for luminous AGN using lower luminosity limits that are in excess of the peak of the distribution, starting at $L_{\mathrm{bol}}\geq 10^{44}$erg/s, in order to probe the regime of uncommonly high SMBH accretion rates for the TNG100-1 simulation, highlighted by the red to yellow dashed lines in Figure \ref{fig:AGNLums}. Figure \ref{fig:AGNLums} shows what fraction of the total parent sample are in each AGN sample, as well as the total number of AGN above each luminosity cutoff. We find that our most inclusive AGN luminosity limit, $L_{\mathrm{bol}} \geq 10^{44}$ erg/s, includes just under 6\% of galaxies from our parent sample (53,482 AGN out of 1,021,226 galaxies from TNG100-1), whereas our most restrictive sample, $L_{\mathrm{bol}} \geq 10^{45.5}$ erg/s, includes 84 AGN making up just 0.01\% of the parent sample.

\section{The incidence of luminous AGN in pairs and post-mergers}
\label{sec:AGNinPairsAndPMs}
\begin{center}
\begin{figure*}
	\includegraphics[width=2.1\columnwidth]{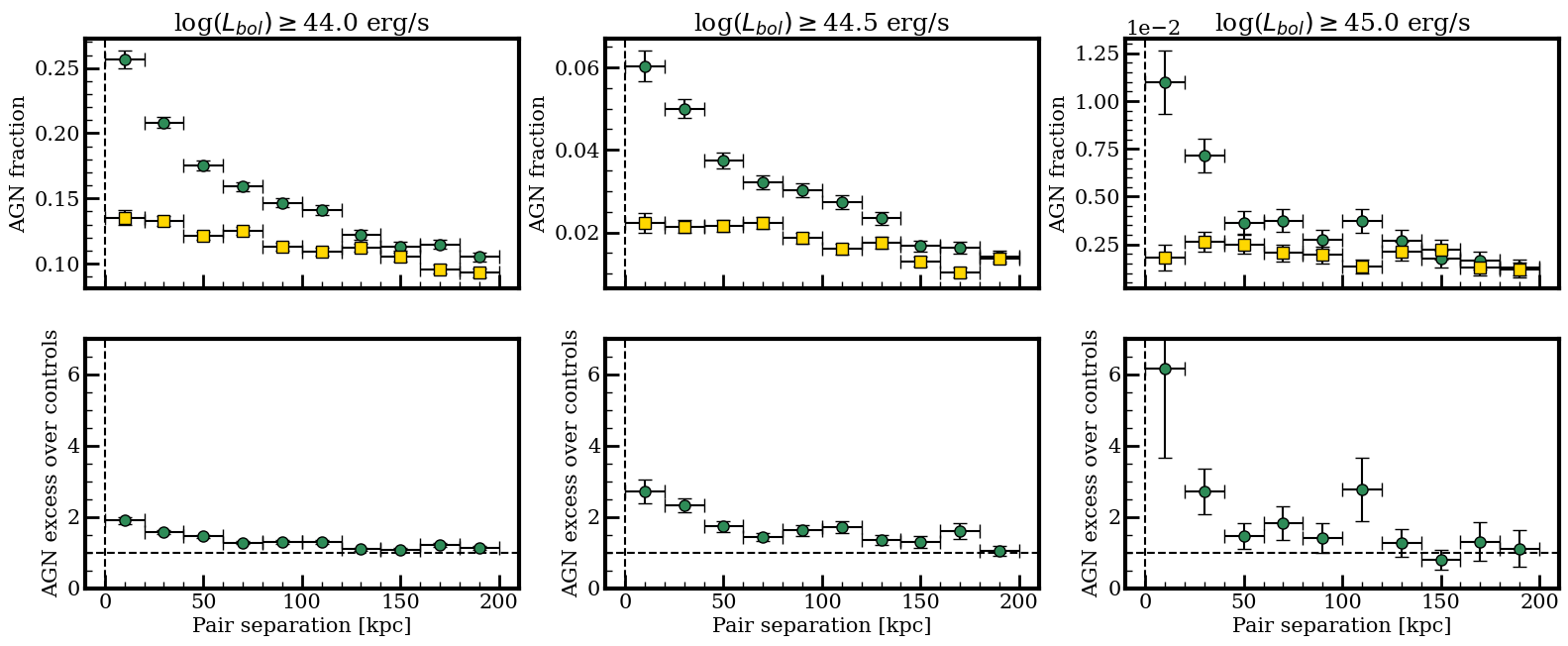}
    \caption{The AGN fraction for galaxies from the pairs sample, and the excess over matched non-pair controls. The top three panels show the fraction of AGN in the pairs (green circles) and the non-pair controls (yellow squares), binned by the pair separation up to 200 kpc. Each top panel, from left to right, shows the luminous AGN fraction for increasing minimum AGN luminosity: $L_{\mathrm{bol}}\geq 10^{44.0}$ erg/s, $L_{\mathrm{bol}}\geq 10^{44.5}$ erg/s, and $L_{\mathrm{bol}}\geq 10^{45.0}$ erg/s. The bottom panels show the fractional excess of AGN in the pairs, equal to $f_{\mathrm{AGN}}^{\mathrm{pairs}}/f_{\mathrm{AGN}}^{\mathrm{controls}}$, for each luminosity threshold. Error bars on the y-axis are calculated by propagating the standard error from binomial statistics. Error bars on the x-axis represent the bin width.}
    \label{fig:PairFracAGNExcess_LumCuts}
\end{figure*} 
\end{center}

In the section that follows, we investigate how commonly pairs and post-mergers host a luminous ($L_{\mathrm{bol}}\geq 10^{44}$ erg/s) AGN. We determine both the fraction of pairs that host a luminous AGN and the excess over non-pair controls in Section \ref{subsec:AGNInPairs}, and extend the analysis to the fraction and excess of luminous AGN along the merger sequence in both the pre and post-coalescence phase in Section \ref{subsec:AGNvsTPMTUM}. We emphasize that for all the results presented here, we calculate the fraction of all paired galaxies which host a luminous AGN. Note that this is not the same as the AGN fraction calculated pairwise (the fraction of pairs hosting one or more AGN).

\subsection{The luminous AGN fraction in pairs}
\label{subsec:AGNInPairs} 

\begin{figure}
    \centering
	\includegraphics[width=\columnwidth]{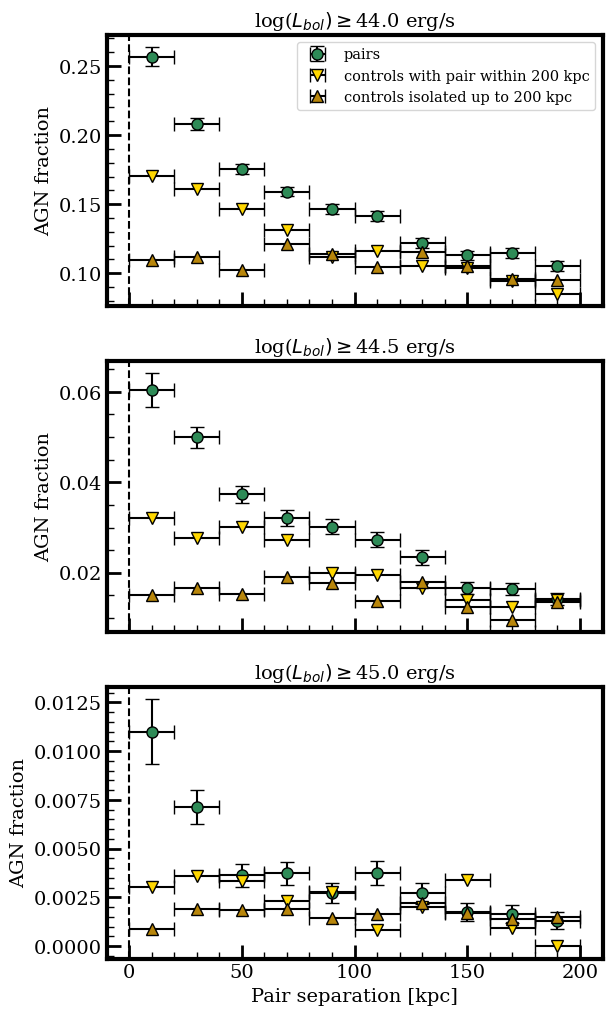}
    \caption{The role of environment in the AGN fraction of controls. The three panels show the fraction of AGN in the pairs (green circles), following the format of Figure \ref{fig:PairFracAGNExcess_LumCuts}, where the panels, from top to bottom, show the luminous AGN fraction for increasing minimum AGN luminosity: $L_{\mathrm{bol}}\geq 10^{44.0}$ erg/s, $L_{\mathrm{bol}}\geq 10^{44.5}$ erg/s, and $L_{\mathrm{bol}}\geq 10^{45.0}$ erg/s. The yellow and orange triangles show the AGN fraction in controls, separated by the distance to their nearest companion, where upside down yellow triangles are controls with a companion within 200 kpc, and the orange triangles are isolated controls. Error bars on the y-axis are calculated by propagating the standard error from binomial statistics. Error bars on the x-axis represent the bin width.}
    \label{fig:PairFracAGNExcess_SepControls}
\end{figure}

To begin, we calculate the fraction of luminous AGN in our total pairs sample and their matched non-pair controls, in bins of pair separation, shown in Figure \ref{fig:PairFracAGNExcess_LumCuts}. The top panels of the figure show the luminous AGN fraction in pairs (green circles) and controls (yellow squares), for pair separations up to 200 kpc. From left to right, the panels of Figure \ref{fig:PairFracAGNExcess_LumCuts} show the fraction of AGN above $10^{44}$ erg/s, $10^{44.5}$ erg/s, and $10^{45}$ erg/s. The bottom panels show the relative excess of AGN in pairs compared with their matched controls, where the range of the y-axis is fixed in order to facilitate comparison between luminosity thresholds. The error in the x-axis represents the minimum and maximum range of the bin width. The y-axis error in the top panel is the binomial error on the AGN fraction in the paired and control galaxy samples. The y-axis error in the bottom panel is the fractional uncertainties for each binned separation of paired and control galaxies, added in quadrature.

In the top three panels of Figure \ref{fig:PairFracAGNExcess_LumCuts}, we consistently find an increasing AGN fraction with decreasing pair separation, and that the highest AGN fraction occurs in the closest pair bin. As expected, we find the highest absolute AGN fractions in the least restrictive AGN luminosity cutoff ($L_{\mathrm{bol}}\geq 10^{44}$erg/s), with an AGN fraction of 26\% in the closest pair separation bin. However, these AGN are also common in the non-pair controls (a 14\% incidence), resulting in a maximum excess over controls of a factor of $\sim 2$ at the closest separations. In the centre and right-most panels, we see that the absolute AGN fractions are decreasing with increasing AGN luminosity, again as expected, with at most 6\% of the closest pairs hosting AGN with luminosity greater than $10^{44.5}$ erg/s and only 1\% hosting AGN with luminosity greater than $10^{45}$ erg/s. However, when comparing to the fraction of luminous AGN in the non-pair controls, we find that these rare AGN are much more commonly in pairs, up to a factor of 3 to 6. The large excess at the closest pair separations found in the work presented here are larger than what has been previously shown in the EAGLE cosmological simulation \citep[][for a lower luminous AGN threshold of $L_{\mathrm{bol}}\geq 10^{43}$erg/s]{McAlpine2020}{}{}, but are consistent with the excess of AGN found in the closest BH-BH pairs from TNG100-1 \citep[][for $\geq 0.7$ Eddington ratio AGN]{Bhowmick2020}{}{}, which further supports the dependence of the AGN excess on AGN luminosity. In addition, we find a particularly strong excess of the most luminous AGN in the paired galaxies with a separation 100-120 kpc. Upon closer investigation, we find that the excess is statistically robust, and reflects an increase in the fraction of AGN found in pairs at $\sim$90-110 kpc separation. We investigate the characteristics of the paired galaxies at $\sim$100 kpc separation, and find that there is a slight excess of very dense environments (10\% more galaxies with more than 10 companions), compared with adjacent bins. Although the excess of companions is weak, it supports a scenario in which the wide separation AGN enhancements may be connected to the large scale environment, possibly through the enhanced frequency of interactions.

In the bottom panels of Figure \ref{fig:PairFracAGNExcess_LumCuts}, a modest yet statistically significant excess of AGN in pairs persists out to separations of 150-200 kpc, for all luminous AGN cutoffs. The modest (a factor of $\sim$ 1-2) AGN excess in pairs is similar to what has been found in the EAGLE simulation \citep[][]{McAlpine2020}{}{} for separations less than 80 kpc, however, our results demonstrate a statistically significant excess out to wider separations than found in both the EAGLE simulation (excess up to 80 kpc \citealt{McAlpine2020}) and in observational studies (excess up to $\sim$ 60-80 kpc in \citealt{Ellison2011}, \citealt{Satyapal2014}, and \citealt{Bickley2023}). While the discrepancy may be alleviated when considering projected separations, as projected separations will always appear to be equal to or smaller than the 3D pair separation, the effect of projection was shown to be minimal in \cite{Patton2020}. An alternative explanation is the role of control matching methodology. Notably, \cite{Ellison2013}, who apply the most similar matching scheme to our fiducial matching scheme from Section \ref{subsec:controls}, find an excess of AGN in pairs at least out to separations of 80 kpc (their upper limit in pair separation). We test whether we find a significant difference in the AGN excess in pairs when applying the control matching scheme of \cite{Ellison2011} and \cite{Bickley2023} (only matching in redshift and stellar mass), and find our results nearly unaffected, where the maximum pair separation with a significant excess of AGN decreases slightly from 150 to 130 kpc. Therefore, the difference of control matching methodology is also insufficient to explain the excess at wide separations. Despite the tension with observations, the enhancement of AGN at wide separations (greater than 100 kpc) is consistent with the enhancement of SFR found in TNG pairs in \cite{Patton2020}. Additionally, we note that the excesses at wide separation are very small (typically smaller than a factor of 1.5), and may not be significant enough to be detected without very large sample statistics. 

Finally, we address the trend of increasing AGN fraction in `non-pair' controls as a function of the separation of their corresponding matched pair. The increasing AGN fraction in controls is due to the environment matching methodology (see Section \ref{subsec:controls}), which allows for the presence of galaxies with a companion in the `non-pair' control sample in order to match the local environment in the pairs. In Figure \ref{fig:PairFracAGNExcess_SepControls}, we once again show the AGN fractions in pairs for three minimum AGN luminosities (following the format of Figure \ref{fig:PairFracAGNExcess_LumCuts}), shown in green circles. However, we now separate the matched controls into those with a companion within 200 kpc (upside down yellow triangles), and those isolated up to 200 kpc (orange triangles). We demonstrate that only the controls with companions show an increasing trend in absolute AGN fraction. We emphasize that we do not attempt to quantify the effect of additional companions on the AGN fraction in the work presented here. We instead allow for additional companions to be present in the control sample in order to control for their effect, and therefore isolate the contribution from only the closest companion.

\subsubsection{AGN fraction in pairs: primaries and secondaries}
\label{subsubsec:primAndSec}

\begin{figure*}
	\includegraphics[width=1.5\columnwidth]{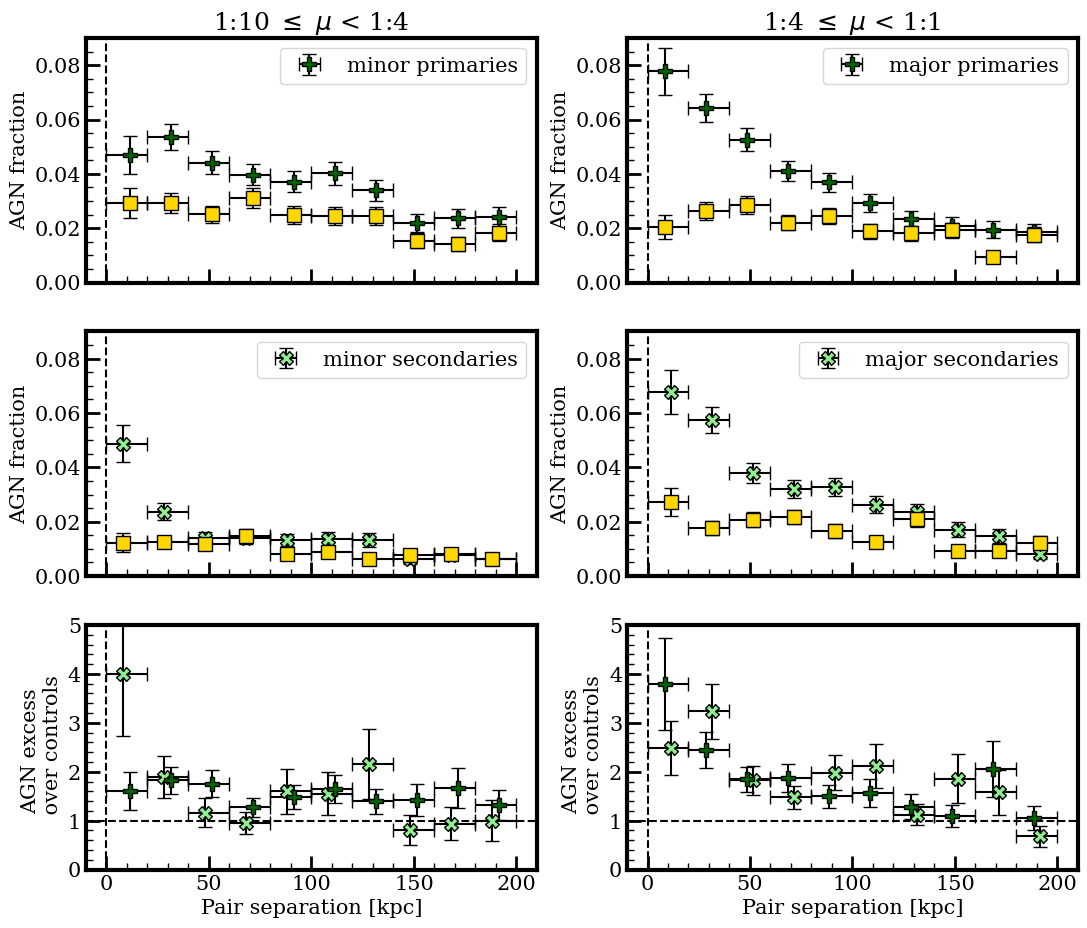}
    \caption{The AGN fraction in pairs, separated into the primary (more massive) and secondary (less massive) galaxies, and separated by minor (1:10<$\mu$<1:4, left) and major ($\mu > 1:4$, right) pairs. The top panels show the fraction of AGN in the primary galaxies (dark green crosses), and the central panels show the fraction of AGN in the secondary galaxies (light green x). Once again, non-pair controls are shown in yellow squares, and the pair galaxies are binned by their pair separation. In all panels, the threshold for luminous AGN is $L_{\mathrm{bol}}\geq 10^{44.5}$ erg/s. The bottom panels show the excess of AGN in the pairs over controls, equal to $f_{\mathrm{AGN}}^{\mathrm{pairs}}/f_{\mathrm{AGN}}^{\mathrm{controls}}$. Error bars on the y-axis are calculated by propagating the standard error from binomial statistics. Error bars on the x-axis represent the bin width.}
    \label{fig:PairFracAGNExcess_MRsep}
\end{figure*}
\begin{center}
\begin{figure*}
	\includegraphics[width=2.1\columnwidth]{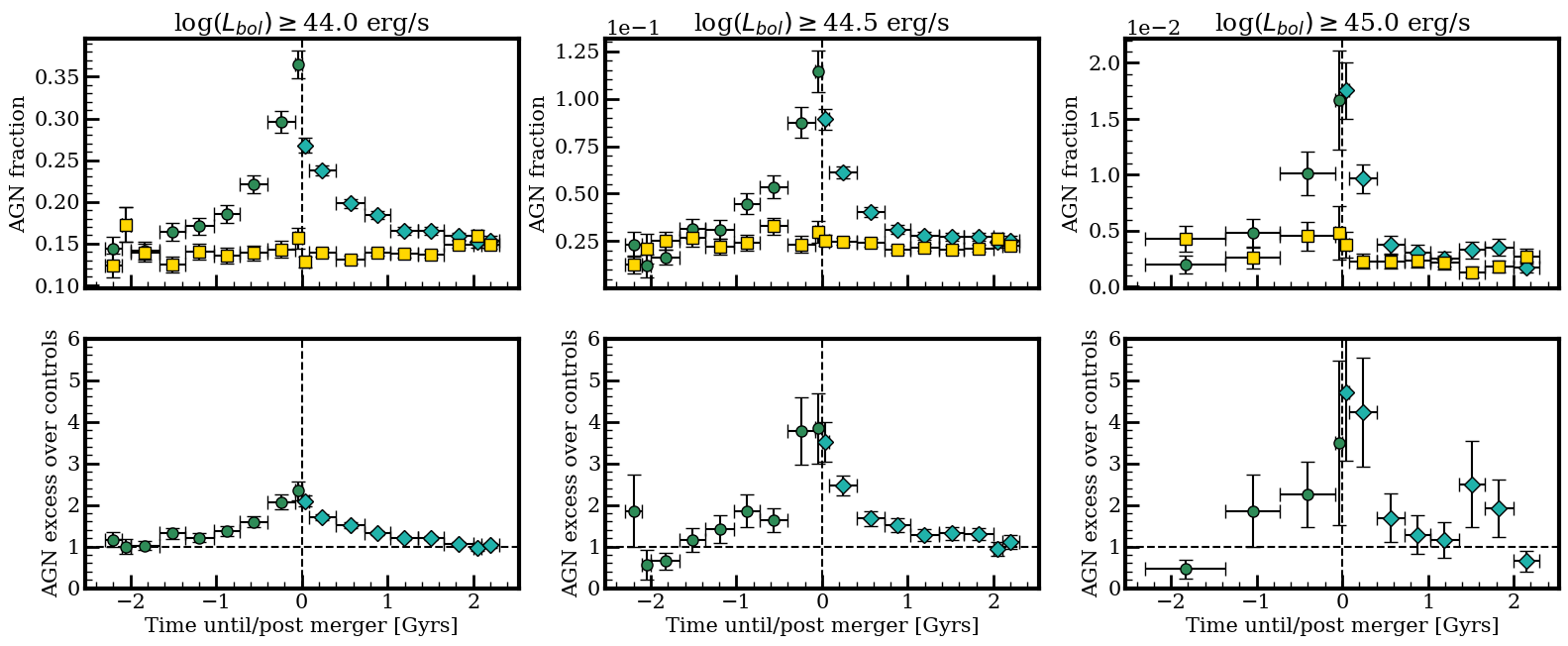}
    \caption{The evolution of the AGN fraction across the merger sequence. The top three panels show the fraction of AGN in the merging pairs (green circles) and post-mergers (teal diamonds) and their controls (yellow squares). Each panel, from left to right, shows the fractions of AGN with increasing minimum AGN luminosity: $L_{\mathrm{bol}}\geq 10^{44.0}$ erg/s, $L_{\mathrm{bol}}\geq 10^{44.5}$ erg/s, and $L_{\mathrm{bol}}\geq 10^{45.0}$ erg/s. The bottom panels show the AGN fractional excess of the pairs or post-mergers equal to $f_{\mathrm{AGN}}^{\text{pairs or PMs}}/f_{\mathrm{AGN}}^{\text{controls}}$, for each luminosity threshold. Error bars on the y-axis are calculated by propagating the standard error from binomial statistics. Error bars on the x-axis represent the bin width. }
    \label{fig:PreMPM_AGNFrac}
\end{figure*}
\end{center}

The strength of gaseous inflows, which are thought to trigger AGN, are influenced by many of the pair properties such as mass ratio, gas fractions, and orbit properties \citep[][]{DiMatteo2007,Lotz2010b,Lotz2010a,Capelo2016}{}. Additionally, the individual galaxies in each pair may react differently to the dynamical interaction, particularly in the most mass disparate cases we consider. We therefore continue our investigation of the AGN fraction in pairs, looking at the difference in the AGN fraction of primary (more massive) and secondary (less massive) galaxies, as well as the role of pair mass ratio. We only consider one luminous AGN cutoff for the experiment, $10^{44.5}$ erg/s, as the experiment requires a strong enough excess signal in order to evaluate the different in triggering between the subcategories while maintaining high enough statistics for a significant interpretation of results. Figure \ref{fig:PairFracAGNExcess_MRsep} shows the AGN fraction separated by mass ratio (minor $1:10<\mu<1:4$ on the left, and major $\mu>1:4$ on the right), and subdivided into primaries (top panel), and secondaries (centre panel). Primary galaxies are shown as dark green crosses, and secondaries as light green x markers. Once again, controls are shown as yellow squares. The bottom panels of the figure show the excess of AGN in pairs over controls.

Focusing first on the major and minor primaries (top panels), we find that in both mass ratio regimes, there is trend of increasing AGN fraction with pair separation. However, we find that the trend is stronger in major pairs. A weaker trend in the minor pairs is consistent with primary galaxies that are less dynamically disturbed by the presence of a low mass companion. 

In the central panels, we investigate the secondary galaxies, in both minor and major pairs. In the minor secondaries, we only find a significant AGN fraction at the closest pair separations (5\% at separations 0-20 kpc), which sharply drops off at separations greater than 20 kpc. Our results therefore indicate that for these minor secondaries, the triggering of luminous AGN only occurs over a very limited regime. The major secondaries show a quantitatively similar, but slightly higher, AGN fraction at the closest pair separations. However, the AGN fraction of secondaries in major pairs does not drop off as sharply with increasing pair separation. 

In the bottom panel we show the excess of AGN over matched controls, in minor (left) and major (right) pairs. Interestingly, the highest excess of AGN over controls occurs in major primaries and minor secondaries (excesses of a factor $\sim $3-5), even though the absolute fraction of AGN is almost twice as high in major primaries than minor secondaries. However, beyond 0-20 kpc, the quantitative excess is similar between all four categories, a factor of 1-2. In addition, every subcategory except minor secondaries demonstrate a statistically significant excess of AGN up to separations of 150-200 kpc. 

Taken together, our results demonstrate that at the closest pair separations, AGN fractions are typically higher in major pairs than in minor pairs, consistent with observational studies that find AGN fractions are highest in major pairs \citep[][]{Silverman2011,Stemo2021}{}{}. However, beyond $\sim $50 kpc, we find that the AGN fractions are similar between the major and minor pairs. In terms of excesses, we similarly find a higher AGN excess in major pairs at the closest pair separations, but a similar factor of enhancement in the major and minor pairs beyond $\sim $50 kpc.

\subsection{AGN fraction throughout the merger sequence}
\label{subsec:AGNvsTPMTUM}

Having established the excess of luminous AGN in pairs, we now investigate how the AGN fraction changes along the merger sequence, starting from the pre-coalescence phase all the way through until long past the merger is complete. In the following experiment, we reduce our pairs sample (157,046 control matched pairs) to those that experience a merger event within the simulation time (7125 merging pairs). We will refer to this subset of pairs as "merging pairs". In contrast to the analysis of pairs in the previous section, which was conducted as a function of projected separation, we can now quantify the time until merger from the simulation merger tree. The pairs that cannot be associated with a future merger event (and are hence dropped from the merging pairs sample) are generally excluded because they have not yet merged by the $z=0$ simulation snapshot. We note that, for this reason, there is a slight bias to remove pairs from the later snapshots of the simulation. However, we additionally note that there is little to no preference for merging pairs to be selected from specific regimes in mass ratio. Once again we compare the AGN fraction in merging pairs with a matched non-pair control sample, using the controls as identified in Section \ref{subsec:controls}. After coalescence, we investigate the incidence of AGN in the post-merger sample compared with non-interacting controls selected using the methodology outlined in Section \ref{subsec:controls}.

The AGN fraction along the merger sequence is shown in Figure \ref{fig:PreMPM_AGNFrac}. The top panels show the AGN fraction, for luminous AGN cutoffs from left to right: $L_{\mathrm{bol}}\geq 10^{44}$ erg/s, $L_{\mathrm{bol}}\geq 10^{44.5}$ erg/s, and $L_{\mathrm{bol}}\geq 10^{45}$ erg/s. The merging pairs are shown as green circles, the post-mergers as teal diamonds. Controls are shown as yellow squares. We additionally apply an adaptive binning scheme, wherein we increase the bin size over which we calculate an AGN fraction if fewer than 3 luminous AGN are found in the merging pairs/post-mergers or controls. The bottom panels show the excess of AGN in the merging pairs or post-mergers over their controls, for each AGN luminosity cutoff. Once again, we fix the extent of the y-axis in order to emphasize the dependence of the AGN excess on the AGN luminosity threshold. 

In the top panels of Figure \ref{fig:PreMPM_AGNFrac}, there is a qualitatively similar trend for all three luminous AGN cutoffs: increasing AGN fraction leading up to coalescence, the highest AGN fractions in the bins closest to coalescence, and a decline in AGN fraction in the post-coalescence phase. As expected, the absolute fraction of AGN decreases with increasing AGN luminosity. Notably, merging pairs in the two temporal bins closest to coalescence have an AGN fraction similar, though slightly higher, than the pairs from Figure \ref{fig:PairFracAGNExcess_LumCuts} between 0-20 kpc, with up to 35\% of pre-coalescence merging pairs hosting luminous AGN with $L_{\mathrm{bol}} \geq 10^{44}$ erg/s. Despite the higher absolute AGN fraction in merging pairs, we find the excess over controls is at most a factor of $\sim$ 2. In comparison, luminous AGN with $L_{\mathrm{bol}} \geq 10^{44.5}$ erg/s and $L_{\mathrm{bol}}\geq10^{45}$ erg/s have higher excesses of AGN over controls right before coalescence, by a factor of 4. We see a similar result in the AGN fractions in post-mergers, with the AGN excess getting stronger with increasing AGN luminosity cutoffs. In addition, for the luminous AGN with $L_{\mathrm{bol}} \geq 10^{44}$ erg/s and $L_{\mathrm{bol}}\geq10^{44.5}$ erg/s, the AGN fraction and AGN excess in merging pairs exceeds the post-mergers, but the trend reverses for $L_{\mathrm{bol}} \geq 10^{45}$ erg/s, where post-mergers have a slightly higher AGN fraction and AGN excess, which may reflect that the highest luminosity AGN need a stronger dynamic interaction than can be achieved in the pair phase to sustain their elevated SMBH accretion rates.

Concerning the timescales presented in Figure \ref{fig:PreMPM_AGNFrac}, we find that for all three luminous AGN cutoffs, the AGN fraction in merging pairs is statistically enhanced beginning around 1-2 Gyr pre-coalescence. The timescale of luminous AGN enhancement in post-mergers is similarly long-lived for $L_{\mathrm{bol}} \geq 10^{44}$ erg/s and $L_{\mathrm{bol}}\geq10^{44.5}$ erg/s, but drops off more rapidly for $L_{\mathrm{bol}} \geq 10^{45}$ erg/s (in-line with controls around 800 Myr). Such a difference may reflect that the most luminous AGN events are difficult to sustain post-coalescence, but may be triggered in the multiple rounds of pericentric interactions that are experienced in merging pairs, yielding an AGN excess earlier in the merger sequence. However, we also note that the sample statistics are particularly small in the regime of the most luminous AGN, and the difference in timescales is also consistent with the higher uncertainties in the luminous AGN fractions. The subtle enhancements in the luminous AGN fractions for $L_{\mathrm{bol}} \geq 10^{44}$ erg/s and $L_{\mathrm{bol}}\geq10^{44.5}$ erg/s persist on timescales slightly shorter than the timescale of enhanced SMBH accretion rates found in \cite{ByrneMamahit2023}, 1-2 Gyr compared with greater than 2 Gyr. However, the SMBH accretion rate enhancement found in \cite{ByrneMamahit2023} did not account for the absolute accretion rate, and therefore would also include the low level enhancement to SMBH accretion rate occurring in low luminosity AGN. 

Overall, our results demonstrate that the enhancement of AGN activity in the pre-coalescence phase is statistically comparable to the enhancements seen in post-mergers, even exceeding post-mergers for some luminous AGN cutoffs. However, in the most luminous AGN, post-mergers demonstrate higher levels of enhancement.

\section{The Interaction Fraction of Luminous AGN}
\label{sec:IntFracOfAGN}

In the previous sections we have demonstrated the luminosity dependent excess of AGN hosted in pairs and post-merger galaxies. We have found that significant AGN excesses persist out to wide separations beyond 100 kpc. We additionally find that, in the most luminous AGN, post-mergers host the strongest excess over their matched controls, a factor of $\sim$3.5-6.5 in recent post-mergers. Despite this clear connection between an interaction and the triggering of a luminous AGN, our analysis has not yet addressed whether a merger is \textit{required} in order to produce such an extreme accretion event.  Indeed, in our previous work (\citealt{ByrneMamahit2023}) we found that most AGN in TNG are not related to recent mergers, supporting previous simulation \citep[][]{Steinborn2018,McAlpine2020}{}{} and observational \citep[][]{Schawinski2011,Schawinski2012,Hewlett2017,Villforth2014}{}{} work.  However, our previous work only assessed the fraction of AGN in post-mergers and did not consider the pre-coalescence regime. With our expanded analysis here that includes both pairs and post-mergers, we can now conduct a more complete assessment of the fraction of AGN in TNG that are triggered by an interaction, either pre- or post-coalescence.

In the section that follows, we investigate the contribution of interactions to the luminous AGN population. In order to robustly identify post-mergers down to a minimum mass ratio of 1:10, we enforce a total stellar mass limit of at least $10^{10} M_{\odot}$ for the entirety of the section. We therefore investigate luminous AGN in the 302,068 galaxies with a total stellar mass above $10^{10} M_{\odot}$.

\subsection{Fraction of Pairs and Post-mergers in Luminous AGN}
\label{subsec:frac_AGN_in_interactions}

\begin{figure}
	\includegraphics[width=\columnwidth]{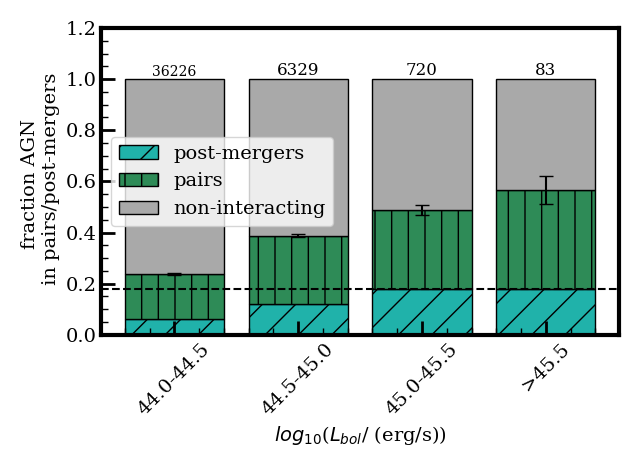}
    \caption{The interaction fraction of luminous AGN, of increasing luminosity in units of $\mathrm{log}_{10}(L_{\mathrm{bol}}/(\mathrm{erg}/\mathrm{s}))$ from left to right. Galaxies in each bin are assigned a classification of either post-merger (within 500 Myrs post-coalescence, shown in teal), pair (within pair separation of 100 kpc, shown in green), or non-interacting (grey). The numerical value above each bar indicates the number of galaxies per bin, and the black horizontal dashed line is the interaction fraction (post-merger + pair fraction) for the entire parent sample (of total stellar mass above $10^{10} M_{\odot}$). The vertical error bar is the fractional error given by binomial statistics.}
    \label{fig:AGNFracIntExcess_bar}
\end{figure}

In order to label each galaxy's interaction status, we assign each galaxy a label of post-merger (coalescence within the last 500 Myrs), pair (< 100 kpc pair separation), or non-interacting. We note that while we do not explicitly include a selection for pre-mergers (galaxies that will merge in the next 500 Myrs), the majority of these cases are captured within the pair selection. In the event a luminous AGN is both a post-merger and a pair, we assign the label post-merger. We note that while the order of priority in which we assign galaxies as pairs or mergers will affect the individual pair and post-merger fractions, the total interaction fraction (pair + post-merger) remains the same.

Figure \ref{fig:AGNFracIntExcess_bar} shows the interaction fraction of galaxies as a function of their AGN luminosity, divided into the contributions from each type of interaction: post-mergers shown in teal and pairs shown in green. On top of each bar is the number of AGN in each bin, and the horizontal dashed line shows the interaction fraction for the total parent sample, i.e. the interaction fraction for non-AGN and AGN combined. The fraction of interacting galaxies in the lowest AGN bin, $10^{44} \leq L_{\mathrm{bol}} <10^{44.5} \mathrm{erg/s}$, is just over the normal incidence of interactions for the parent sample. However, as we move towards the higher and rarer AGN luminosities, the fraction of AGN in interactions increases. At the highest luminosity bin,$L_{\mathrm{bol}} \geq 10^{45.5}$ erg/s, we find that 55\% of the AGN sample are in pairs or post-mergers, which contribute 35\% and 20\% respectively. In every AGN bin, we find that the pair fraction exceeds the post-merger fraction, as is expected given that post-mergers (within 500 Myrs of coalescence) are less common than pairs (within 100 kpc) in our parent sample.

To summarize, our main findings regarding the incidence of interactions in luminous AGN are: post-mergers and pairs play an increasingly important role in AGN triggering with increasing AGN luminosity, however not all luminous AGN are triggered by interactions, even at the highest luminosities. In addition, we find that the contribution to the interaction fraction from pairs exceeds the post-mergers. We note that the contribution from pairs still outnumber post-mergers, even when we relax the criterion for post-merger classification up to 1 Gyr post-coalescence. 

\subsubsection{Excess of interactions in AGN compared with non-AGN controls}
\label{subsubsec:controlMatchAGN}

In Figure \ref{fig:AGNFracIntExcess_bar}, we quantified the fraction of pairs and post-mergers in AGN. However, it is common in observations to match AGN to non-AGN controls in order to compare the incidence of interactions between samples without the influence of the unique population characteristics in the luminous AGN \citep[][]{Koss2010,Cisternas2011,Kocevski2012,Bohm2013,Kocevski2015,Rosario2015,Mechtley2016,Villforth2017,Ellison2019,Marian2019,Marian2020}{}{}. We therefore test whether the increasing fraction of interactions in the luminous AGN is robust in comparison with matched controls.

For each AGN from each luminosity bin in Figure \ref{fig:AGNFracIntExcess_bar}, we identify a single best-matched non-AGN control from a control pool of galaxies with AGN luminosities below $10^{43}$ erg/s. Recall that $L_{\mathrm{bol}} \sim 10^{43.2}$ erg/s was the most common luminosity from our parent sample of TNG100-1 galaxies (see Figure \ref{fig:AGNLums}). We therefore select non-AGNs with luminosities below this threshold to select for galaxies in a relatively `inactive' phase, in comparison to the standard set by all the TNG100-1 galaxies meeting our general criteria. We additionally test matching to non-AGN controls with lower bolometric luminosities, $L_{\mathrm{bol}}\leq 10^{42}$ erg/s, and find our results qualitatively unchanged. 

For the work presented here, we apply a simple matching criteria commonly used in observations, matching in redshift and stellar mass \citep[][]{Kocevski2012,Mechtley2016,Villforth2017,Ellison2019,Marian2019,Marian2020}{}{}. From the control pool, we select a single control galaxy that is matched exactly in redshift and is the closest match in total stellar mass. If a stellar mass match cannot be found to within 0.3 dex, we discard the luminous AGN from the sample. We successfully match 90\% of the AGN with $10^{44}\leq L_{\mathrm{bol}}< 10^{44.5}$ erg/s, 99\% of AGN with $10^{44.5}\leq L_{\mathrm{bol}}< 10^{45}$ erg/s, and 100\% of AGN with $L_{\mathrm{bol}}\geq 10^{45}$ erg/s.

\begin{figure}
	\includegraphics[width=\columnwidth]{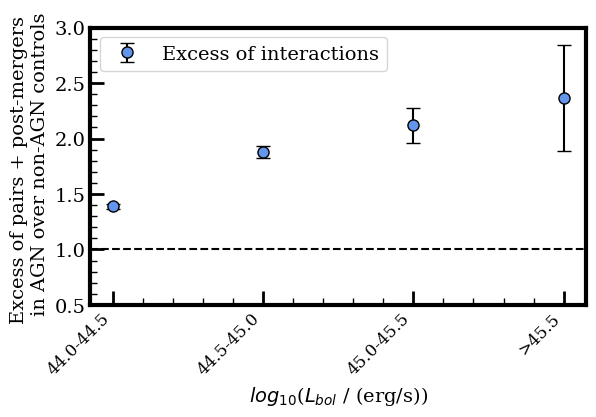}
    \caption{The excess in the interaction fraction of luminous AGN compared with matched non-AGN controls. The interaction fraction is defined as the fraction of galaxies in a pair within 100 kpc separation or a post-merger within 500 Myr of coalescence. Control galaxies are matched exactly in redshift, and selected as the best match in total stellar mass from a control pool with $L_{\mathrm{bol}} < 10^{43}$ erg/s. Error bars on the y-axis are calculated by propagating the standard error from binomial statistics.}
    \label{fig:AGNExcessOverControls}
\end{figure}

We show the excess of interactions in AGN in Figure \ref{fig:AGNExcessOverControls}. The y-axis is the fractional excess of interactions (recent post-merger within 500 Myrs + pairs within 100 kpc) in the AGN compared to the matched non-AGN controls. We find a statistically significant excess of interactions in all bins of luminous AGN. The excess of interactions is luminosity dependent, and reaches a maximum $\sim $2.3 in the highest luminosity bin. In comparison, in Figure \ref{fig:AGNFracIntExcess_bar}, we find that the interaction fraction in the most luminous AGN is $\sim $55\%, compared with $\sim $20\% in the parent sample (an excess of $\sim$2.8, without matching to control non-AGN). We therefore find that the interaction fraction is twice as large in the most luminous AGN compared with controls matched in redshift and total stellar mass. 

\section{Discussion}
\label{sec:discussion}

In the following section, we discuss whether the results of the work presented here are broadly in agreement or disagreement with the observational literature on interacting galaxies and AGN. We further investigate sources of tension, such as the dependence of the interaction fraction of AGN on additional properties such as redshift, or how the interaction fraction in AGN is sensitive to our definitions of pairs and post-mergers.

\subsection{Comparison to observations}
\label{subsec:observations}

In the work presented here, we combine the pair and post-merger phase to investigate AGN triggering in interacting galaxies in TNG. In the following section, we comment on how the results of this work compare both qualitatively and quantitatively to the observational literature.

We first presented the analysis of AGN triggering in pairs in Figure \ref{fig:PairFracAGNExcess_LumCuts}, finding that the fraction of luminous AGN, and the excess over controls, peaks at the closest separations (2-3 for $L_{\mathrm{bol}}>10^{44-44.5}$ erg/s, and up to 6 for $L_{\mathrm{bol}}>10^{45}$ erg/s). We additionally find that pairs with wide separations host an excess of luminous AGN by of a factor of 1-2. Some observations find a quantitatively similar maximum excess, between 2-3, of optical \citep[][]{Ellison2011,Ellison2019,Bickley2023}{}{} and X-ray AGN \citep[][]{Silverman2011}{}{} in pairs, although we note disagreements with observational studies that find a lower excess for closely separated pairs \citep[][]{Steffen2023}{}{} or those that do not find an AGN excess in pairs \citep[][]{Scott2014,He2023}{}{}. Additionally, the excess we report for the highest luminosity AGN is compatible with excesses found in obscured systems \citep[][]{Satyapal2014,Weston2017,Goulding2018,Ellison2019,Bickley2023}{}{}.

Despite an encouraging agreement in the maximum AGN excess in the closest separated pairs, we find that AGN excesses extend to pair separations far exceeding those seen in observations ($\sim $ 60-80 kpc in \citealt{Ellison2011}, \citealt{Satyapal2014} and \citealt{Bickley2023}, compared with 150 kpc in the work presented here). Such a discrepancy cannot be explained with projected separations or differences of control matching methodology. However, we note that the small excesses found at wide pair separations may require large sample sizes in order to statistically distinguish the AGN fractions in pairs and controls. For example, for the most luminous AGN ($L_{\mathrm{bol}}\geq 10^{45}$ erg/s), the AGN fraction of wide pairs (100 kpc separation) is between 0.2 and 0.4 percent, or 2-4 AGN per 1000 paired galaxies.

In our investigation of AGN triggering through the merger sequence (Figure \ref{fig:PreMPM_AGNFrac}), we reported the highest AGN excess (a factor of 3.5-6.5) for AGN with luminosity greater than $10^{45}$ erg/s in recent post-mergers (within 500 Myrs of coalescence). However, for AGN at luminosities of $L_{\mathrm{bol}}>10^{44-44.5}$ erg/s, we find a higher excess in the merging pairs immediately pre-coalescence. Observations have found that AGN triggering is strongest at the coalescence phase compared with close pairs \citep[][]{Satyapal2014,Ellison2019,Bickley2023}{}{}, in agreement with our highest luminosity AGN but in contrast to our intermediate luminous AGN. 

In Figure \ref{fig:AGNFracIntExcess_bar}, we re-framed our investigation towards the contribution of interactions to the luminous AGN population, and we demonstrated that there is a dominant population of pairs+post-mergers in the highest bin ($L_{\mathrm{bol}}\geq 10^{45.5}$ erg/s) of luminous AGN, and a luminosity dependence in the interaction fraction. Our results therefore agree with observations finding that the most luminous AGN host a high fraction of interactions (\citealt{Bennert2008},\citealt{Urrutia2008},\citealt{RamosAlmeida2011},\citealt{Bessiere2012},\citealt{Chiaberge2015} for radio loud AGN,\citealt{Glikman2015},\citealt{Hong2015},\citealt{Kocevski2015},\citealt{Fan2016},\citealt{Ellison2019}) and a luminosity dependence in the merger fraction in AGN \citep[][]{Treister2012,Fan2016,Glikman2015,Pierce2022}{}{}, but in disagreement with studies that find either a subdominant fraction of interactions in AGN (\citealt{Koss2010},\citealt{Cisternas2011},\citealt{Kocevski2012},\citealt{Bohm2013},\citealt{Villforth2014},\citealt{Chiaberge2015} for radio quiet AGN,\citealt{Mechtley2016},\citealt{Hewlett2017},\citealt{Donley2018},\citealt{Marian2019},\citealt{Villforth2017},\citealt{Villforth2019},\citealt{Marian2020}), or no luminosity dependence in the merger fraction in AGN \citep[][]{Villforth2014,Villforth2023}{}{}. It is important to note that although we find a dominant (more than 50\%) interaction fraction in the most luminous AGN, it is not an overwhelming majority, suggesting that even in the highest luminosity regime, secular triggering processes still play a significant role. We discuss this point further in Section \ref{subsec:NonInt}.

Finally, we have investigated whether the high interaction fraction in luminous AGN persists when compared with redshift and stellar mass matched controls, shown in Figure \ref{fig:AGNExcessOverControls}. We find that all luminous AGN thresholds ($L_{\mathrm{bol}}\geq 10^{44,44.5,45}$ erg/s) have a statistically significant excess of interactions, and that the interaction excess grows with increasing AGN luminosity. Our results are in agreement with observations that find an excess of interactions over non-AGN controls \citep[][]{Koss2010,Kocevski2015,Ellison2019,Marian2020}{}{}, and are quantitatively similar to the excesses found in \cite{Ellison2019} and \cite{Kocevski2015} (factors 2-3), but lower than the excesses in \cite{Koss2010} and \cite{Marian2020}.

\subsection{Breaking down the excess in the interaction fraction of AGN}
\label{subsec:IntFracvsX}

We note in Section \ref{subsec:observations} that the results of the work presented here are in conflict with observational studies which do not find a dominant interaction fraction in AGN \citep[e.g.][ and others]{Koss2010,Cisternas2011,Kocevski2012}{}{}. One major caveat in the comparison of our results and observations is whether the luminous AGN sample we study in the work presented here is comparable to those in observations. The results we present above are for the luminous AGN sample in aggregate, which may have a very different underlying distribution from those in observations due to numerous factors such as, for example, AGN observability \citep[][]{Menzel2016}{}{}. While matching the AGN characteristics between our sample and those in observations is beyond the scope of this work, in the following section, we present how the interaction fraction in AGN varies as a function of a few key parameters: redshift, large scale environment, and host properties (gas fraction and SFR), in order to discuss whether the pair and post-merger fractions vary significantly for AGN occupying different regimes of these parameters. Once again, here we only consider one luminosity definition for our luminous AGN sample, $L_{\mathrm{bol}}\geq 10^{44.5}$ erg/s.

\begin{figure*}
\subfloat[\label{subfig:excess_vs_z}]{\includegraphics[width = 3in]{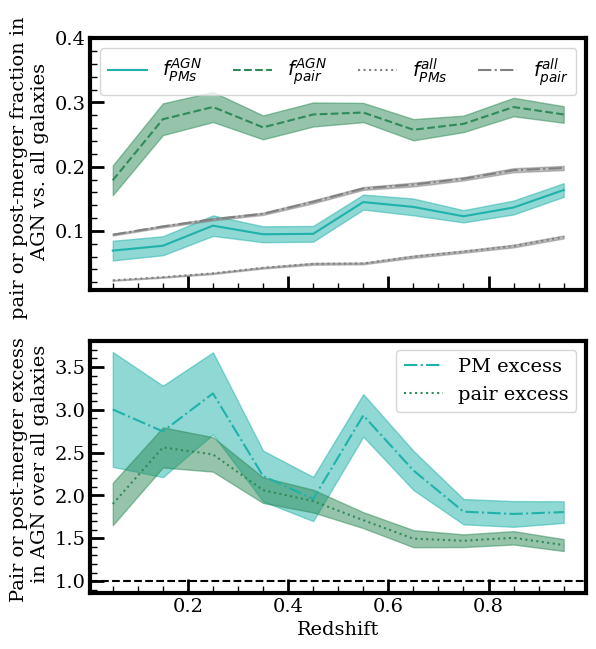}} 
\subfloat[\label{subfig:excess_vs_N2}]{\includegraphics[width = 3in]{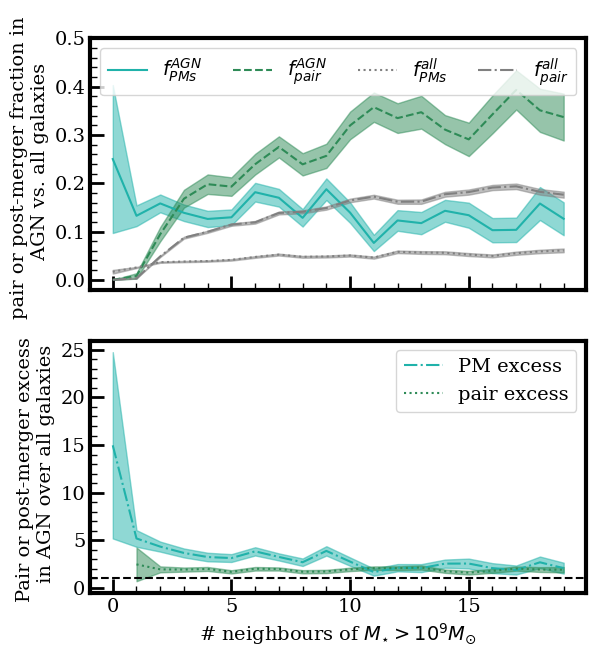}}\\
\subfloat[\label{subfig:excess_vs_gas}]{\includegraphics[width = 3in]{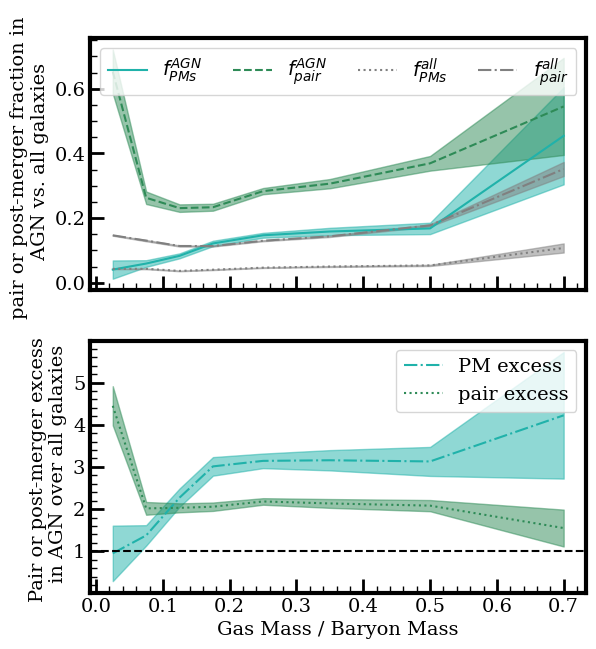}}
\subfloat[\label{subfig:excess_vs_SFR}]{\includegraphics[width = 3in]{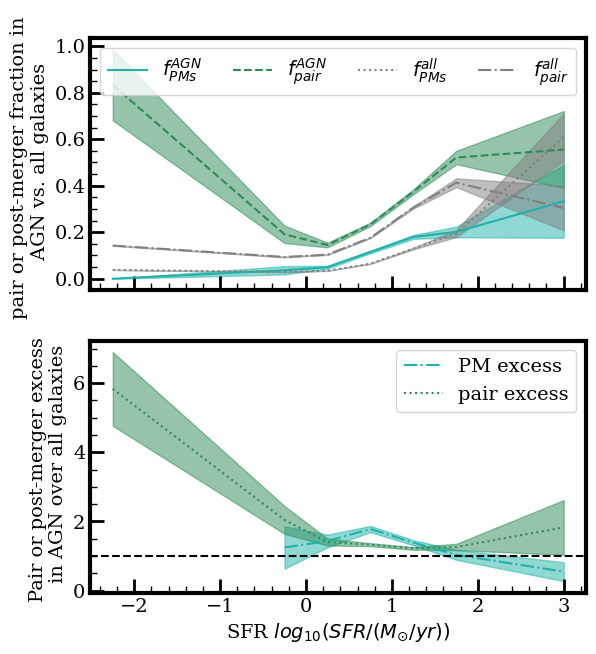}} 
\caption{The dependence of the pair and post-merger fractions of AGN on: (a) Redshift, (b) the number of neighbours above $10^{9} M_{\odot}$ within 2 Mpc, (c) the gas fraction (gas mass over total baryonic mass, in twice the half mass radius), and (d) star formation rate (in twice the half mass radius). For each figure, the top panel shows the fraction of pairs in the luminous AGN (defined as $L_{\mathrm{bol}}\geq 10^{44.5}$ erg/s) in the green dashed line and the fraction of post-mergers in the luminous AGN in the solid teal line. The pair and post-merger fractions in the total parent sample are shown in the grey dashdot and dotted lines respectively. In the bottom panel of each figure, the excess (relative to the total parent sample) is shown for post-mergers (shown in teal) and pairs (shown in green). All y-errors are calculated by propagating the standard error from binomial statistics.}
\label{fig:dependenceofIntFrac}
\end{figure*}

In Figure \ref{fig:dependenceofIntFrac}, we break down the luminous AGN by four different parameters: redshift (Figure \ref{subfig:excess_vs_z}), the number of neighbours above a mass $10^{9} M_{\odot}$ within 2 Mpc (Figure \ref{subfig:excess_vs_N2}), the gas fraction (gas mass over total baryonic mass, in twice the half mass radius, Figure \ref{subfig:excess_vs_gas}), and star formation rate (in twice the half mass radius, Figure \ref{subfig:excess_vs_SFR}). In the top panel of each figure, we show the fraction of pairs in the luminous AGN in the green dashed line, and the fraction of post-mergers in the luminous AGN in the solid teal line. The pair and post-merger fractions in the total parent sample (all TNG100-1 galaxies with a total stellar mass greater than $10^{10} M_{\odot}$ and $z<1$) are shown in the grey dashdot and dotted lines respectively. In the bottom panel of each figure, we show the excess of pairs or post-mergers in AGN compared with the total parent sample. We apply an adaptive binning in each parameter, allowing the bin size to increase such that each point has a minimum count of at least 3 pairs or post-mergers.

Beginning with redshift, observational studies suggest that the merger-AGN connection may vary between low and high redshift, as high-redshift galaxies, which have higher gas fractions, may achieve optimal AGN triggering conditions without the need for merger events \citep[][]{Rosario2015,Marian2020}{}{}. In the top panel of Figure \ref{subfig:excess_vs_z}, we see that the fraction of pairs and post-mergers in both the AGN and total parent sample, is increasing from $z=0$ to $z=1$, which reflects the known evolution in the galaxy merger rate with redshift \citep[][]{Lin2004,Lin2008,Rodriguez-Gomez2015,Man2016,Conselice2022}{}{}. If we only consider the $z=0$ bin, there is a 30\% interaction fraction in AGN ($\sim$ 10\% post-mergers + 20\% pairs) which is subdominant, but in excess of the interaction fraction in the parent sample ($\sim $10\%). In the bottom panel of Figure \ref{subfig:excess_vs_z}, there is a clear trend of increasing pair excess down to redshift 0.1 (with a slight decrease in excess in the lowest redshift bin). There is a similar trend shown in post-mergers, with more scatter (expected due to the smaller statistics of the post-merger sample). The lowest excess of post-mergers occurs at z=0.8 and the highest excess at z=0-0.2. Our results therefore agree with the prediction that galaxy interactions play an increasingly important role in AGN triggering at low redshift. In fact, we note that redshift offers a possible explanation for the disagreement with \cite{Koss2010} and \cite{Marian2020}, both of which investigate AGN at low redshift and find a higher excess over non-AGN controls than is presented here. In comparison, the most luminous AGN sample in the work presented here, $L_{\mathrm{bol}}\geq 10^{45}$ erg/s, has an average redshift of 0.67. 

Similarly to redshift, observational studies demonstrate that AGN triggering excesses may be sensitive to differences in environment, suggesting that AGN fractions decline in the highest density environments \citep[][]{Khabiboulline2014,Li2019,Duplancic2021}{}{}. We investigate whether the interaction fraction of AGN is sensitive to an increasing density of environment in Figure \ref{subfig:excess_vs_N2}. When separated by the number of neighbours above a mass of $10^{9} M_{\odot}$, the post-merger and pair fractions in both the AGN and parent samples evolve differently. The pair fraction of both the AGN and the parent sample generally increases with increasing number of neighbours, which is expected as it is more likely to have a nearby companion in denser environments. In contrast, we see that the post-merger fraction is highest (25\%) in isolated systems. Such systems of isolated AGN are rare, however where they do occur, a quarter of these AGN are post-mergers. From the bottom panel of the figure, we see that the increase of pair fraction in both the AGN and the parent sample is consistent, producing a relatively flat trend in the pair excess with the number of neighbours. Similarly, beyond the regime of isolated AGN, the post-merger excess is relatively stable, although somewhat lower for AGN with more than 10 companions. The persistent excess in both the pair and post-merger fraction in AGN with increasingly higher density environments suggest that interactions continue to contribute an excess to AGN triggering even in dense environments, which is in agreement with the observations of AGN in pairs vs groups by \cite{Duplancic2021}.

In Figures \ref{subfig:excess_vs_gas} and \ref{subfig:excess_vs_SFR}, we investigate how the pair and post-merger fractions depend on the gas fraction (gas mass over baryon mass in twice the half mass radius) and star formation rate (in twice the half mass radius). Simulations \citep[][]{DiMatteo2007}{}{} and observations \citep[][]{Woods2007,Scudder2015,Cao2016}{}{} have investigated the correlations between gas content, SFR, morphology, and color on the strength of AGN triggering in galaxy interactions, with conflicting results. For example, in \cite{ByrneMamahit2023} we find the stronger SMBH accretion rate enhancements in gas rich post-mergers, whereas \cite{DiMatteo2007} do not find that the initial gas content in interacting pairs strongly influences the strength of the inflow/starburst. We therefore examine whether the pair and post-merger fractions in AGN are sensitive to the gas-content and SFR of the AGN host. We note that for both the gas fraction and SFR, we apply an adaptive binning and allow the bins to widen if fewer than 3 pairs or post-mergers are found in the AGN or parent sample, in order to calculate a meaningful excess.

Beginning with gas fraction in Figure \ref{subfig:excess_vs_gas}, we demonstrate that in the lowest gas fraction bin, 60\% of gas poor AGN are in pairs. The pair fraction drops as we move to slightly higher gas fractions (dropping to $\sim$20\%), and for gas fraction bins $\geq 0.1$, the pair fraction is increasing. A somewhat opposite trend occurs in post-mergers, which are the least common in the most gas poor AGN and become increasingly common with a maximum post-merger fraction of 40\% in the most gas rich AGN. Considering the pair and post-merger fractions in the total parent sample (grey), we see that the post-merger fraction is relatively flat with gas fraction, if not subtly enhanced at the highest fractions. The pair fraction of the total sample follows a similar trend to the pair fraction of AGN. In the bottom panel of the figure, the excess of AGN in pairs peaks at the lowest gas fraction bin, with an excess of $\sim$ 4.5. Beyond the lowest gas fraction bin, the excess stabilizes around 2 for increasing gas fraction. In comparison, gas poor AGN demonstrate no excess of post-mergers, whereas gas rich AGN have the highest excess of post-mergers, between 2.5-5.5. Our results suggest that pairs and post-mergers play important roles in triggering AGN in different regimes of AGN host gas fraction. The highest excess of post-mergers in gas rich AGN hosts is complementary to the results of \cite{ByrneMamahit2023}, that gas rich post-mergers are most likely to host AGN. However, the enhanced pair fraction in gas poor AGN hosts is less complementary to the simulations and observations quoted above. A possible explanation may be that the gas poor pairs hosting AGN may have a gas rich companion which is providing fuel for the enhanced AGN activity, whereas gas poor post-mergers may be remnants of mergers between two similarly gas poor galaxies which are less likely to trigger a luminous AGN event. However, such a scenario is not supported by \cite{Cao2016} who find that interactions between spiral and elliptical galaxies demonstrate no significant enhancements in SFR, which would suggest little inflow of gas to the galaxy centre. Such scenarios are also less likely to affect the study of luminous AGN, which are preferentially found in gas-rich and star forming hosts \citep[][]{Rosario20131,Ellison20192,Jarvis2020,Shangguan2020,Koss2021}{}{}.

Finally, Figure \ref{subfig:excess_vs_SFR} investigates the post-merger and pair incidence as a function of star formation rate. We note that in the figure, the lowest SFR bin includes TNG100-1 galaxies with unresolved SFRs. In the top panel, for SFRs greater than 1 $M_{\odot}/\mathrm{yr}$, the fraction of luminous AGN host galaxies undergoing interactions in post-mergers or pairs increases with star formation rate. There is a similar increase in the post-merger and pair fractions of the parent sample, where the fraction of galaxies at the highest SFRs in interactions,luminous AGN or not, approaches 40-60\%. A high fraction of interactions in high SFR AGN hosts, but relatively modest fraction of interactions at intermediate SFRs, agrees with simulations which demonstrate that merger driven AGN and SFR enhancements are rarely temporally correlated due to the stochastic nature of SMBH accretion rates \citep[][]{Hickox2014,Volonteri20151}{}{} and is in agreement with observations that find a connection between AGN and SFR enhancements in only the most highly star-forming galaxies \citep[][]{Schweitzer2006,Lutz2010,Shao2010,Santini2012,Bernhard2016,Xie2021}{}{}. In the bottom panel of Figure \ref{subfig:excess_vs_SFR}, high SFR AGN have only a modest excess, of a factor 1-2, of pairs and post-mergers around SFR$\sim 1-10 M_{\odot}/yr$, and no statistically significant excess of pairs or post-mergers at the highest SFRs. The lack of pair or post-merger excess at the highest SFRs reflects the preference for AGN to reside in star forming galaxies \citep[][]{Schweitzer2006,Lutz2010,Shao2010,Santini2012,Bernhard2016,Xie2021}{}{}. Since the grey line traces the behaviour of the total parent sample, and the incidence of AGN increase with increasing SFR, the grey line becomes more associated with the AGN population. For the lowest SFRs, despite the rarity of luminous AGN in quenched galaxies, a high fraction of the luminous AGN are in pairs, $\sim$ 80\%, compared with less than 20\% in the total parent sample, yielding a very high pair excess of $\sim$ 5-7. The strong excess of pairs suggest that pairs play an important role in triggering luminous AGN, particularly in gas poor and quenched galaxies. We further investigate the AGN in gas poor and quenched pairs, and find that in quenched AGN pairs, the companion galaxies have gas fractions between 10-25\%, whereas quenched non-AGN pairs have companions with similarly low gas fractions below 10\%. Our results would therefore agree with the scenario proposed above that AGN in gas poor and quenched pairs are fueled by gas from a companion.

\subsection{Non-Interacting Luminous AGN}
\label{subsec:NonInt}

In Figure \ref{fig:AGNFracIntExcess_bar}, we found that in the most luminous AGN sample, 55\% of the AGN can be associated to either post-merger or pair interactions. However, while 55\% is a dominant contribution, we consider the possibility that the remaining 45\% of `non-interacting' luminous AGN lie just outside our bounds for selecting post-mergers and pairs. 

From the most luminous AGN bin, we select the 35/83 galaxies that do not fall into the category of recent (within 500 Myr of coalescence) post-mergers or pairs within 100 kpc separation. We find that of the 35 galaxies, 7 have pair separations within 200 kpc. Specifically, two of these wider pairs will merge within the next 200 Myrs. The remaining 5 wide pairs are not going to merge within at least 1 Gyr. In addition, we find that of the remaining 28 galaxies, 6 are post-mergers within 1 Gyr of coalescence (but more than 500 Myr past coalescence). Therefore, even if consider a broader sample of `interacting' galaxies (including wider pairs and late stage post-mergers), we find at most 70\% of the most luminous AGN to be associated with interactions. 

Since even our most luminous AGN bin still has a significant fraction of non-interacting hosts, our findings suggest that secular processes must be capable of triggering even the most luminous AGN in the simulation. However, we mention a few final caveats to the discussion above. The first numerical consideration concerns the application of an Eddington rate maximum to the SMBH accretion rates in TNG (see Section \ref{subsec:IllustrisTNG}). One may imagine that interactions could trigger even more luminous AGN than are allowed within the bounds of Eddington rate limited accretion, and that the SMBH accretion rates fueled by interactions are numerically suppressed due to the TNG model. It is therefore possible that the interaction fraction observed in our most luminous AGN is lower than what may be expected without an Eddington limited accretion rate model. On the other hand, the anchoring of SMBHs to the centre of the potential well will allow for optimal SMBH accretion rates, since the gas density will be highest where the potential is deepest, and therefore SMBH accretion rates in TNG could be overestimated due to the numerical methods. Finally, it is possible the remaining `non-interacting' cases lie just outside the boundaries of our mass criterion for pairs and post-mergers, specifically just below the minimum mass ratio threshold of 1:10. The role of such minor mergers and interactions (mass ratios below 1:10) in AGN triggering remains relatively unexplored, and represents an intriguing avenue for future works.

\section{Conclusions}
\label{sec:conclusion}

In the work presented here, we study AGN triggering in pair and post-merger galaxies. We investigate the connection between the most luminous AGN ($L_{\mathrm{bol}}\geq 10^{44}$ erg/s) and interacting galaxies from the IllustrisTNG simulation. Our three key results are summarized in the following points:

\begin{itemize}
    \item \textbf{Interactions trigger AGN in the pair phase.} AGN with luminosities $L_{\mathrm{bol}}>10^{44}$ erg/s are more commonly found in pairs when compared to matched non-pair controls. In addition, pairs demonstrate a statistically significant excess of AGN over controls, which persists up to separations of 150 kpc (Figure \ref{fig:PairFracAGNExcess_LumCuts}). We find a luminosity dependence in the excess of AGN over controls, with the highest excess in the closest pairs ranging from a factor of 2 (for $L_{\mathrm{bol}}>10^{44}$ erg/s) to a factor of 6 ($L_{\mathrm{bol}}>10^{45}$ erg/s).
    
    \medskip
    
    \item \textbf{Interactions trigger AGN 1.5 Gyr before and after coalescence.} We investigate the evolution of the luminous AGN fraction throughout the merger sequence, finding 1) the number of high luminosity AGN is enhanced up to 1.5 Gyr before and after coalescence, although slightly shorter-lived for the most luminous AGN (Figure \ref{fig:PreMPM_AGNFrac}) and 2) the strength of the AGN excess over matched controls is luminosity dependent, with the highest excess in AGN of luminosities $>10^{45}$ erg/s in immediate post-merger galaxies (factor of 3-7). 
    
    \medskip
    
    \item \textbf{The interaction fraction of AGN is luminosity dependent.} In Figure \ref{fig:AGNFracIntExcess_bar}, we demonstrate that the total fraction of interacting galaxies (pairs within 100 kpc separation + post-mergers within 500 Myr of coalescence) in the AGN sample increases with AGN luminosity, up to a maximum of 55\% for AGN with $L_{\mathrm{bol}}>10^{45.5}$ erg/s. We emphasize that on their own, neither pairs nor post-mergers make up a dominant fraction (35\% pairs and 20\% post-mergers). In addition, we apply a control matching methodology to compare the luminous AGN to non-AGN controls, and find a statistically significant excess of interactions in all luminous AGN bins (Figure \ref{fig:AGNExcessOverControls}), and the highest excess for the highest AGN luminosity
    
    \medskip
    
In addition to these three main results, we investigate a number of further demographic dependencies for the luminous AGN-merger connection. In Section \ref{subsec:IntFracvsX}, we find the lowest absolute fraction of post-merger and pairs in low redshift AGN, but the highest excess over the parent sample. Similarly, we demonstrate that while isolated AGN are rare, they host a post-merger fraction much higher (5-20 times higher) than what is expected for the typical isolated galaxy. Finally, we report that 60 and 80\% of gas poor and quenched AGN hosts are found in pairs, compared with a 20\% pair incidence in typical gas poor and quenched galaxies. In contrast, the most gas rich AGN hosts are more commonly associated with post-mergers, with 4 times more post-mergers in the gas rich AGN than in typical gas rich galaxies, suggesting that pairs and post-mergers may contribute to AGN triggering differently depending on the characteristics of the AGN host.

In conclusion, in the research presented here we have demonstrated the importance of the pair phase in triggering the most luminous AGN, and demonstrated that interactions (pair phase and post-mergers combined) are associated with more than 50\% of the most luminous AGN in TNG100-1. Despite this, a significant fraction (30-45\%) of the most luminous AGN are not associated with interactions, as identified by our methodology, which suggests a significant contribution of secular AGN triggering events or a possible contribution from minor interactions below the minimum mass ratio explored in our work ($\mu$<1:10).
    
\end{itemize}

\section*{Acknowledgements}
We acknowledge and thank the IllustrisTNG collaboration for providing public access to data from the TNG simulations. SJB acknowledges the receipt of the Dr. Margaret Perkins Hess Research Fellowship from the University of Victoria and SJB and SW acknowledge graduate fellowships from the Natural Sciences and Engineering Research Council of Canada (NSERC);  Cette recherche a été financée par le Conseil de recherches en sciences naturelles et en génie du Canada (CRSNG). This research was enabled in part by support provided by WestGrid (www.westgrid.ca) and the Digital Research Alliance of Canada (alliancecan.ca). SLE and DRP gratefully acknowledge the NSERC of Canada for Discovery Grants which helped to fund this research.

\section*{Data Availability}
The data used in this work are publicly available at \hyperlink{}{https://www.tng-project.org}.



\bibliographystyle{mnras}
\bibliography{paperFinal}









\bsp	
\label{lastpage}
\end{document}